\documentclass[journal,twocolumn,10pt,letter]{IEEEtran}

\IEEEoverridecommandlockouts
\ifCLASSINFOpdf
\else
\fi

\usepackage{bbold}
\usepackage{amsfonts}
\usepackage{nonfloat}
\usepackage{amssymb}
\usepackage[cmex10]{amsmath}
\usepackage{amsthm}
\usepackage{amssymb}
\usepackage{mathrsfs}
\usepackage{amsbsy}
\usepackage{setspace}
\usepackage{graphicx}
\usepackage{newclude}
\usepackage{latexsym}
\usepackage{lettrine} % I added this packagehttps://www.overleaf.com/project/5caf4bdce350a8305ec7a022
\usepackage{flushend}

\usepackage{authblk}
\usepackage{subfigure}

\usepackage{scalerel,stackengine}
\stackMath

\usepackage{amsthm}

\newtheoremstyle{named}{}{}{\itshape}{}{\bfseries}{.}{.5em}{\thmnote{#3's }#1}
\theoremstyle{named}

\usepackage{amsthm}

\theoremstyle{definition}

\usepackage[table]{xcolor}
\usepackage{multirow}
\usepackage{rotating}
\usepackage{booktabs}
\usepackage{bbm} % indicator function
\usepackage{amsmath,epsfig,cite,amsfonts,psfrag}
\usepackage{lipsum}
\usepackage{cuted}
\usepackage{color}

\usepackage{epstopdf}
\usepackage{float}
\usepackage{mathtools}
\usepackage{bbm}
\usepackage{atbegshi}% http://ctan.org/pkg/atbegshi
\usepackage{balance}
\usepackage[utf8]{inputenc}
\usepackage{multirow}

\renewcommand{\tablename}{Algorithm}

\usepackage{accents}
\newlength{\dhatheight}

\allowdisplaybreaks

\graphicspath{{figures/}}
\allowdisplaybreaks

\makeatletter
\floatstyle{plain}
\newfloat{twocolequfloat}{b}{zzz}
\floatname{twocolequfloat}{Equation}

\makeatother

\begin{document}
%\title{ A Blind RLNC-Based Transmission Policy for Low-Feedback and Blocked Multiple Access Networks }

%\title{An Overhearing-Driven NOMA Scheme to Minimize Latency in Low-Feedback Networks}

%\title{An Overhearing-Driven NOMA Scheme for Low-Latency Broadcast in Low-Feedback Networks}

%\title{An Overhearing-Driven NOMA Scheme for Low-Latency Multi-User Transmission in Low-Feedback Networks}

%\title{\fontsize{23pt}{29pt}\selectfont Overhearing-Driven NOMA Scheme for Low-Latency Multi-User Transmission under Limited Feedback}

%\title{\fontsize{21.5pt}{29pt}\selectfont Overhearing-Driven NOMA Transmission for Low-Latency Multi-User Networks with Limited Feedback}

%\title{\fontsize{25pt}{29pt}\selectfont Overhearing-Driven NOMA Transmission for Low-Latency Networks with Limited Feedback}

\title{\fontsize{25pt}{29pt}\selectfont Latency Decoupling in Low-Feedback Multi-User Networks via Overhearing-Driven NOMA}


\author{Mohsen~Abedi,
        Ahmed~Badawy, and Amr~Mohamed
        % <-this % stops a space
\thanks{M. Abedi, A. Badawy, and A. Mohamed are with the Dept. Computer Engineering, Qatar University, Doha, Qatar. Email: \{mohsen.abedi;\,badawy;\,amrm\}@qu.edu.qa.} \vspace{-0.4cm}% <-this % stops a space
}

\maketitle

\thispagestyle{empty}  % Remove page number from first page only


\begin{abstract}
Conventional wireless protocols such as Hybrid Automatic Repeat Request (HARQ) rely on frequent and timely feedback, which becomes impractical in low-feedback regimes including non-terrestrial networks and massive IoT. This limitation is particularly critical in heterogeneous multi-user systems with unknown and asymmetric channels, where a single weak user can dominate the overall completion latency. We propose Overhearing-driven Non-Orthogonal Multiple Access (ONOMA), a novel cross-layer transmission scheme that minimizes latency without requiring instantaneous or statistical CSI at the transmitter. ONOMA integrates Random Linear Network Coding (RLNC) with symbol-aware NOMA and explicitly exploits overhearing and acknowledgment timing. In the first phase, users overhear RLNC transmissions, and the relative timing of acknowledgments is used to implicitly infer channel strength ordering. In the second phase, symbol reconstruction enables interference-free decoding for strong users, effectively decoupling user latencies. An adaptive power allocation policy is derived from acknowledgment timing-based channel estimates. Analytical and simulation results show that ONOMA outperforms TDMA, multicast, FDMA, inter-session, and classical NOMA, reducing completion time by up to $34\%$ in two-user and $50\%$ in larger asymmetric networks.
\end{abstract}


% \fi
\IEEEpeerreviewmaketitle

\section{Introduction}

%\textcolor{blue}{Th role of feedback (HARQ, aknowledge):}

In modern wireless communication systems spanning 3G, 4G, 5G, and Wi-Fi networks, timely and reliable feedback is critical for efficient data delivery, high spectral efficiency, and low latency. Hybrid Automatic Repeat Request (HARQ) mechanisms rely on frequent acknowledgement (ACK/NACK) exchanges to support link adaptation, error correction, and throughput optimization across these standards \cite{mehaseb2015classification}. For example, in 4G LTE and 5G New Radio (NR), HARQ employs sub-millisecond feedback loops to enable ultra-reliable low-latency communication (URLLC) and enhanced mobile broadband (eMBB) services \cite{series2015imt}. Similarly, IEEE 802.11 uses block ACK and frame-level acknowledgements, often combined with frame aggregation, to enhance reliability under dynamic channel conditions \cite{liu2025wifi7}. However, in emerging scenarios such as non-terrestrial networks (NTNs), massive Internet of Things (IoT) deployments, and remote healthcare monitoring, frequent feedback is impractical due to long delays, limited uplink resources, or energy constraints \cite{tataria20216g,chowdhury20206g,dang2020should}. As a result, conventional HARQ-based protocols degrade in low-feedback regimes, motivating transmission strategies that reduce feedback dependence while preserving low latency and high reliability.

%\textcolor{blue}{Low feedback networks or highly variant channels:}

Rapid channel variation severely degrades HARQ performance, as the channel state information (CSI) used for link adaptation becomes outdated by the time retransmissions occur, weakening the feedback mechanism \cite{andrews2014will}. Fast channel fluctuations also render ACK/NACK feedback and modulation and coding scheme (MCS) selections obsolete, resulting in repeated decoding failures and reduced throughput \cite{goktepe2023distributed}. Consequently, in high-mobility or fast-fading environments such as vehicular networks and unmanned aerial vehicle (UAV) communications, conventional HARQ suffers from excessive latency and poor reliability, motivating feedback-reduced transmission strategies. To address this, advanced protocols that minimize per-packet acknowledgments have been developed. For instance, Fountain codes support reliable broadcast with limited feedback \cite{nguyen2023rateless}, while uplink scenarios with strict latency constraints, such as 5G NR URLLC, adopt grant-free Non-Orthogonal
Multiple Access (NOMA) and lightweight IoT random access mechanisms \cite{shahab2020grant,palattella2016internet,faridi2010comprehensive}.

Furthermore, operating in high-mobility and non-stationary environments, such as UAV-assisted industrial IoT, necessitates robust link maintenance strategies with limited CSI. Systems in these domains commonly integrate predictive channel estimation techniques, such as Kalman filtering \cite{simon2013iterative}, complemented by physical-layer diversity and multiplexing schemes. These include Space-Time Block Coding (STBC) for transmit diversity \cite{bacsar2010space}, Frequency Hopping Spread Spectrum (FHSS) for interference resilience \cite{rocha2025lrfhss}, and massive Multiple-Input Multiple-Output (MIMO) architectures leveraging statistical CSI for beamforming gain \cite{bjornson2015massive}. Ultimately, the frontier of low-feedback communication is being reshaped by data-driven intelligence, where Reinforcement Learning (RL) and deep neural networks enable autonomous protocol optimization and channel-agnostic encoding, eliminating the reliance on explicit feedback mechanisms \cite{wu2025drl}.

%\textcolor{blue}{RLNC}:

Random Linear Network Coding (RLNC) is a promising  packet coding method that eliminates the need for frequent acknowledgments and retransmissions of specific lost packets~\cite{ho2006rlnc}. It does this by encoding data into random linear combinations of the original packets. A receiver can then recover the data once it collects enough linearly independent combinations, resulting in an efficient and robust form of loss resilience.
Building on this foundation, recent work has optimized RLNC for low-latency communication. In this context, authors in \cite{su2022completion} study minimizing the completion time for decoding RLNC-encoded packets with the assistance of a full-duplex relay. Related works evaluate the latency benefits of RLNC-enabled data planes in cloud-native testbeds \cite{lhamo2025flexnc} and further reduce end-to-end delay and packet loss by strategically placing coded redundancy at optimal overlay nodes, enabling distributed protection across multi-hop paths \cite{xu2025cost}.

While standard protocols such as HARQ are highly effective with timely and reliable feedback, their reliance on frequent acknowledgments for retransmission and rate adaptation becomes a key limitation in low-feedback scenarios. In NTNs, large-scale IoT deployments, and high-mobility vehicular communications, feedback links are constrained by long propagation delays, limited uplink resources, or rapid channel variations that render CSI obsolete. These challenges are exacerbated in multi-user systems with unknown and highly asymmetric channels, where transmission continues until all users decode their packet blocks. As a result, the overall completion time is dominated by the weakest user, leaving strong users idle after early decoding. This latency coupling is a fundamental limitation of conventional schemes such as Time Division Multiple Access (TDMA), multicasting, Frequency Division Multiple Access (FDMA), inter-session network coding, and classical NOMA, motivating the need for transmission strategies that decouple user latencies and ensure low-latency delivery under stringent feedback constraints.

In this work, we propose a novel cross-layer transmission scheme designed to minimize completion latency in low-feedback multi-user networks with asymmetric user channels, where the transmitter has no knowledge of the CSI or the channel statistics. The core mechanism integrates RLNC with a symbol-aware NOMA strategy.
This scheme operates in distinct phases, leveraging passive overhearing to create a strategic advantage. These phases are as follows:

\begin{itemize}
    \item In the initial phase, the transmitter sends RLNC-coded packets to a target user. Other users overhear this transmission, and those with stronger channels decode the data first and send early acknowledgments. This feedback instantly reveals the identities the strongest (first-to-acknowledge) users to the transmitter.
    \item In the second phase, the transmitter generates RLNC-coded packets for both the weak (target user at first phase) and the identified strong user and superimposes their channel-encoded symbols for transmission. This allows the  the strong user to regenerate the weak user's symbols and cancel them from its received signal.
    \item Based on the network state inferred from the acknowledgments during second phase, the transmitter dynamically determines the protocol for the final phase.

\end{itemize} 

By allowing non-target users to overhear RLNC-coded transmissions, the scheme creates strategic side information that prevents strong users from remaining idle. The transmitter exploits acknowledgment timing to implicitly infer channel conditions and adapt power allocation without explicit CSI. In the second phase, symbol reconstruction at the strong user enables interference-free decoding, replacing conventional decoding-based SIC with reconstruction-based subtraction. This decouples user latencies, preventing the weakest user from dominating completion time and achieving substantial latency reduction compared to baseline schemes.

The design philosophy of ONOMA fundamentally differs from classical power-domain multiplexing. Unlike conventional NOMA, where simultaneous transmission relies on channel disparity and successive interference cancellation, ONOMA first creates asymmetric side information through overhearing and then exploits acknowledgment timing to infer relative channel strength without explicit CSI. The resulting superposition is driven by strategically constructed symbol knowledge at the strong user, enabling reconstruction-based interference subtraction. Thus, ONOMA is a latency-oriented protocol that integrates overhearing, feedback timing, and coding to decouple user completion times, rather than merely combining RLNC with NOMA.

Finally, to validate the superiority of the ONOMA scheme, we derive and compare the expected completion times with conventional baselines. The analysis shows that in multi-user networks with asymmetric channels, baseline latency is dominated by the weakest users. In contrast, the proposed scheme mitigates this bottleneck, achieving a lower and more stable completion time.

The rest of this paper is structured as follows: Section II outlines the system model and the foundation for the study. Section III provides the problem statement and the baseline transmission schemes.  Section IV describes the proposed transmission scheme and Section V presents the detailed
simulation results. Finally, Section VI concludes the paper.

%\textcolor{blue}{problem statement}:

%\textcolor{blue}{Contribution of this paper}:

\section{System model}
Consider a point-to-point (PtP) communication link in which the transmitter sends data to a user over a Rayleigh fading channel.  Then, the received signal is modeled as
\begin{equation}
    y=hx+n,
\end{equation}where 
$x$ denotes the transmitted symbol with $\mathbb{E}[|x|^2]=P$, $h \sim \mathcal{CN}(0,\Omega)$ is the complex channel coefficient whose envelope follows a Rayleigh distribution, $n \sim \mathcal{CN}(0,\sigma^2)$ refers to the white Gaussian noise, and $P$ represents the total transmit power.  Here, the channel is assumed  block-fading, and thus, the channel coefficient is considered unchanged during a packet transmission.  In this channel, we assume capacity-achieving coding and sufficiently long packets; hence, the packet loss probability can be approximated by the outage probability, which is derived as
\begin{equation}
    P_{\text{loss}}= \text{Pr}\big(\dfrac{|h|^2 P}{\sigma^2}<\tau(R)\big), \label{eq:packet_loss}
\end{equation}where $\tau(R)=2^R-1$ in which $R=R_c\log _2M$ denotes the target rate, the integer $M$ represents the modulation order and $R_c\in [0,1]$ denotes the code rate.  %So, a larger modulation order and code rate results in a lager target rate, and therefore, a larger packet loss probability. 

\begin{figure*}
    \centering
    \includegraphics[width=0.75\linewidth]{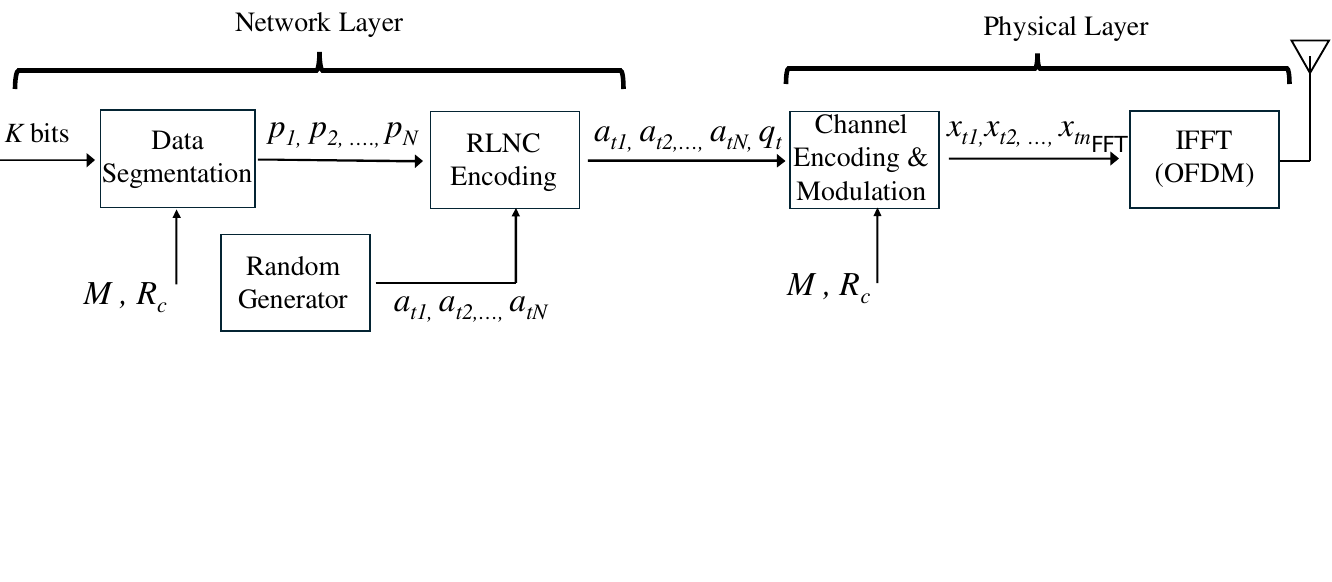}
    \vspace{-2.7cm}
    \caption{Building blocks of a RLNC-based PtP transmitter.}
    \vspace{-0.4cm}
    \label{fig:blocks}
    
\end{figure*}

Since the envelope of  $h$
 follows a Rayleigh distribution, the packet success probability in \eqref{eq:packet_loss} follows an exponential form and is given by
\begin{equation}
P_{\text{succ}} = 1 - P_{\text{loss}}
= \Pr\big(|h|^2 > \frac{\tau(R)\sigma^2}{P}\big)
= \exp\big(\frac{-\tau(R)\sigma^2}{\Omega P}\big),
\label{eq:PtP_P_loss}
\end{equation}
which implies that a smaller target rate and a larger channel gain variance result in a higher packet success probability.

In case of symbol superposition, the received signal at the user is modeled as
\begin{equation}
y = h(\sqrt{\lambda}x + \sqrt{1-\lambda}s) + n,
\end{equation}
where $x$ and $s$ denote the transmitted symbols intended for two different users, with
$\mathbb{E}[|x|^2]=\mathbb{E}[|s|^2]=P$ and 
$\lambda \in [0,1]$ denoting the power allocation coefficient.  Here, the user decodes 
$x$ by treating 
$s$ as an interfering signal.
Then, the packet success probability is formulated as 
\begin{equation}
\begin{aligned}
P_{\text{succ}}
&= \Pr\!\left(\frac{\lambda |h|^2 P}{(1-\lambda)|h|^2 P + \sigma^2} > \tau(R) \right)
 \\
&=  \exp\!\left(-\frac{\tau \sigma^2}{\Omega \big(\lambda - \tau(R) (1-\lambda)\big)P}\right).
\end{aligned}
\label{eq:P_loss_Mac}
\end{equation}
 From (\ref{eq:P_loss_Mac}), the power allocation coefficient should satisfy
 \begin{equation}
 \dfrac{\lambda}{1-\lambda}>\tau(R),
 \label{eq:lambda-requirement}
 \end{equation}to ensure a non-zero packet success probability. %Here, by setting $\lambda=1$, the packet loss probabilities in  (\ref{eq:PtP_P_loss}) and (\ref{eq:P_loss_Mac}) become equivalents.

%.... HARQ process but low feedback 

%.... bad channel condition and blockage
\subsection{Low-feedback systems}
In many communication scenarios, acknowledgment feedback is costly due to long round-trip delays and uplink scarcity, as in NTNs, or infeasible under strict power constraints, as in IoT and healthcare sensing applications. Consequently, receivers return ACK/NACK feedback at longer intervals rather than per packet. We assume each packet is transmitted in a single time slot and that the receiver can return an ACK at least once every $N$ time slots, where $N \gg 1$. The transmitter thus sends a block of $N$ packets $p_1, p_2, \ldots, p_N$ over $\mathrm{GF}(2)$, and the user acknowledges successful decoding of the entire block. Let $T$ denote the number of transmissions (time slots) required for correct reception of the block.

\subsection{Uncoded packet transmission }

In the uncoded transmission scheme, the transmitter sends packets
$p_1, p_2, \ldots, p_N$ sequentially. If any packet is lost, no acknowledgment is returned, and
the transmitter restarts transmission from the first packet due to the lack of loss index information.
The receiver buffers successfully decoded packets across cycles and returns a block-level ACK
upon decoding the last missing packet, with at most one ACK every $N$ time slots. Upon receiving
this ACK, the transmitter terminates the transmission of the packet block.

With this transmission scheme, the cumulative distribution function (CDF) of the completion time $T$ is given by
\begin{equation}
\Pr\!\big(T \le N(k-1)+m\big)
= \big(1-P_{\text{loss}}^{k}\big)^{m}\big(1-P_{\text{loss}}^{k-1}\big)^{N-m},
\label{eq:CDF_uncoded}
\end{equation}
where $k \ge 1$ denotes the number of transmission cycles and $m \in \{1,\ldots,N\}$ denotes the number of transmissions in the last cycle. The, using the tail-sum formula, the expected number of transmissions is
\begin{equation}
\mathbb{E}[T^{\text{uncoded}}]
= \sum_{k=1}^{\infty}\sum_{m=1}^{N}
\Big[1-\big(1-P_{\text{loss}}^{k}\big)^{m}\big(1-P_{\text{loss}}^{k-1}\big)^{N-m}\Big],
\label{eq:ET_uncoded}
\end{equation}
which converges rapidly for any $0 \le P_{\text{loss}} < 1$.
\subsection{ Random Linear Network Coding (RLNC)}
In RLNC-based PtP packet transmission, the network-encoded packet in time slot~$t$ is formed as a linear combination of the $N$ packets over $\mathrm{GF}(2)$, where the combination reduces to a bitwise exclusive-or (XOR) operation
represented as 
\begin{equation}
\begin{aligned}
q_t=\bigoplus_{i=1}^N a_{ti}p_i,
\end{aligned}
\label{eq:RLNC-packet}
\end{equation}
where  $a_{ti}=\{0,1\}$ has a symmetric Bernoulli distribution. Further, the network-encoded packet  $q_t$ has the same size as any individual packet $p_i$, for $1 \leq i \leq N$.

Fig.~\ref{fig:blocks} illustrates the main components of the RLNC-based PtP transmitter at time slot~$t$. At the network layer, $K$ information bits are segmented into $N$ packets $p_1, \ldots, p_N$, which are RLNC-encoded according to~\eqref{eq:RLNC-packet} using the random coding vector $a_{t1}, \ldots, a_{tN}$. The network-encoded packet $q_t$ and its coding vector are then forwarded to the physical layer for channel encoding and modulation at rate $R = R_c \log_2 M$. The resulting symbols are mapped onto OFDM subcarriers and transformed to the time domain via an $n_{\text{FFT}}$-point IFFT, yielding one OFDM symbol corresponding to $q_t$.

\begin{figure}
    \centering
    \includegraphics[width=0.75\linewidth]{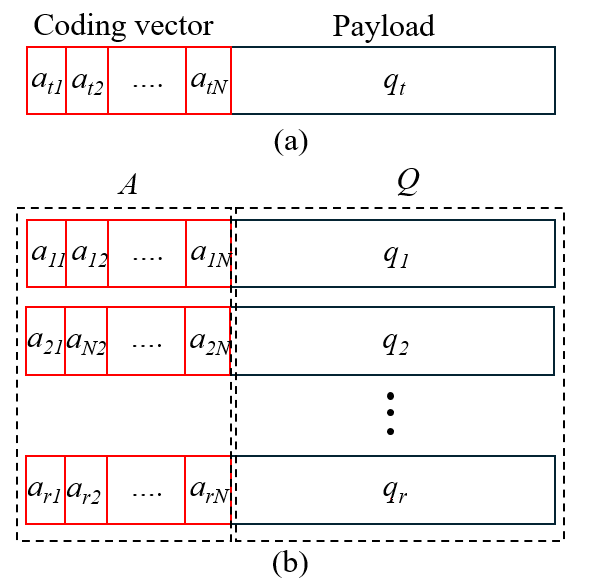}
    \vspace{-0.4cm}
    \caption{A RLNC-based (network-encoded) packet structure; a) Packet structure at time slot $t$ and b) the decoding matrix at the user.}
    \vspace{-0.4cm}
    \label{fig:packet}
\end{figure}

Fig.~\ref{fig:packet}a) shows the structure of an RLNC-encoded packet at time slot~$t$, including the coding vector elements $a_{t1}, \ldots, a_{tN}$. After demodulation and channel decoding, the user checks whether the packet is successfully received. If so, the coding vector is appended to the decoding matrix $A$ and the decoded packet to $Q$, as depicted in Fig.~\ref{fig:packet}b). Once $\mathrm{rank}(A)=N$, the receiver solves $A^{-1}Q$ to recover $p_1, \ldots, p_N$ and sends an ACK, terminating the current block transmission.

In this setting, the probability that the rank of $A$ increases from $r$ to $r+1$ equals the probability that the packet is successfully received and that its coding vector is linearly independent of the previously collected $r$ vectors, given by
\begin{equation}
\begin{aligned}
s_r=P_\text{succ}(1-2^{r-N}),~~~~r=0,1, ..., N-1
\end{aligned}
\label{eq:P_r_to_r+1}
\end{equation}
which shows the fact that as $r$ grows, the probability of incrementing $r$ with a successful packet transmission decreases. Therefore, the expected total number of packet transmissions coincides with the expected number of transmissions until the rank of $A$ reaches $N$ as 
\begin{equation}
\begin{aligned}
\mathbb{E}[T^{\text{RLNC}}]=\sum_{r=0}^{N-1}\dfrac{1}{s_r}=\dfrac{F(N)}{P_{\text{succ}}},
\end{aligned}
\label{eq:ET_RLNC}
\end{equation} where
\begin{equation*}
\begin{aligned}
F(N)=\sum_{r=0}^{N-1}\dfrac{1}{1-2^{r-N}}\approx N+1.6.
\end{aligned}
\end{equation*} 
For RLNC over $\mathrm{GF}(2)$, the expected rank-deficiency overhead converges to the constant
$\sum_{k=1}^{\infty}(2^{k}-1)^{-1}\approx 1.6$, yielding $F(N)\approx N+1.6$ for moderate and large $N$. Hence, when $P_{\text{succ}}$ is sufficiently close to one, the expected number of packet transmissions is approximately $N + 1.6$ which is significantly smaller than that of the uncoded transmission scheme in~\eqref{eq:ET_uncoded}.

\section{Problem statement and the baseline transmission schemes}
Consider a multi-user network in which a transmitter is assigned to send independent (user-specific) blocks of $K$-bits data stream to each of $U$ users, indexed by $u = 1, 2, \ldots, U$.
At the beginning of each transmission block, the transmitter partitions the $K$-bits data stream intended for user $u$ into packets 
$p_1^{(u)}, p_2^{(u)}, \ldots, p_{N^{(u)}}^{(u)}$, for $u = 1,2,\ldots,U$. Here, 
the number of packets for each user at the output of the data segmentation block is given by
\begin{equation}
N^{(u)} = \left\lceil \frac{K}{R^{(u)} n_{\text{FFT}}} \right\rceil=\left\lceil \frac{\bar{K}}{R^{(u)}} \right\rceil,
\label{eq:Nu}
\end{equation}where $R^{(u)} =  R_c^{(u)} \log_2 M^{(u)}$ denotes the transmission data rate of user~$u$, where $M^{(u)}$ represents the modulation order and $R_c^{(u)}$ denotes the assigned channel coding rate. The parameter $n_{\text{FFT}}$ indicates the FFT size at the output of the channel encoder, which corresponds to the number of useful OFDM subcarriers. 
Since $K$ and $n_{\text{FFT}}$ are fixed system parameters in our model, we define the normalized block size as $\bar{K} = K / n_{\text{FFT}}$. Furthermore, the selected data rate $R^{(u)}$ is assumed to remain constant over the entire transmission block.

After partitioning the data streams intended for each user into packets, the transmitter applies an RLNC-based packet transmission scheme in accordance with the selected transmission scheme to generate a network-encoded packet in each time slot. Specifically, for each user $u$, the packets 
$p_1^{(u)}, p_2^{(u)}, \ldots, p_{N^{(u)}}^{(u)}$ are linearly combined at the network layer to form an RLNC-encoded packet, which is subsequently forwarded to the physical layer.
At the physical layer, the network-encoded packet is channel-encoded and modulated with the coding rate $R_c^{(u)}$ and modulation order $M^{(u)}$ assigned to user~$u$, and transmitted over the wireless channel according to the adopted multiple-access strategy. Upon successful reception of a packet, user~$u$ recovers its own channel-decoded packet and appends the associated coding vector to its decoding matrix. User~$u$ returns an ACK to the transmitter once the rank of its decoding matrix reaches $N^{(u)}$, at which point it can successfully recover its entire packet set 
$p_1^{(u)}, p_2^{(u)}, \ldots, p_{N^{(u)}}^{(u)}$.

The objective is to minimize the completion time required to deliver all $K$ information bits to each user, or equivalently, to successfully deliver all $N^{(u)}$ packets $p_1^{(u)}, p_2^{(u)}, \ldots, p_{N^{(u)}}^{(u)}$ for all $u = 1, 2, \ldots, U$.
Without loss of generality, we consider a two-user channel ($U = 2$), in which the transmitter is tasked with delivering a total of $N^{(1)} + N^{(2)}$ packets. The channels of both users are assumed to experience independent Rayleigh fading, modeled as $h^{(u)} \sim \mathcal{CN}(0, \Omega^{(u)})$ for $u = 1, 2$. The noise powers at the users are assumed to be identical, i.e., $\sigma_1^2 = \sigma_2^2 = \sigma^2 = N_0 B$, where $N_0$ denotes the noise power spectral density and $B$ represents the available system bandwidth.

We analyze completion time of two-user networks where at least one user experiences unfavorable channel conditions and thus a low average channel gain, e.g., due to interference, severe weather, burst errors,or blockage. Also, the transmitter is assumed to have no knowledge of the instantaneous CSI or the channel statistics of any user.% We analyze the expected completion time of a two-user network under a set of baseline transmission schemes.

\subsection{RLNC-based TDMA}
In RLNC-based TDMA transmission scheme, the network-encoded packet at the transmitter is generated  as
\begin{equation}
\begin{aligned}
q_{t}^{(u)}=\bigoplus_{i=1}^{N^{(u)}} a_{ti}^{(u)}p_{i}^{(u)},
\end{aligned} ~u=1,2.
\label{eq:RLNC-packet-
TDMA}
\end{equation}
 for each user independently at each time slot, where  $a_{ti}^u=\{0,1\}$ has a symmetric Bernoulli distribution. This packet is then channel-encoded and modulated at the physical layer with transmission rate $R^{(u)}$ and transmitted over the channel using the full available power $P$. As soon as the rank of the decoding matrix of user $u$ reaches $N^{(u)}$, it returns an ACK to the transmitter, upon which the transmitter immediately proceeds with the transmission to the other user.

In this setting, the successful packet probability at each time slot  follows the PtP channel model as in (\ref{eq:PtP_P_loss}) and written as
\begin{equation}
\begin{aligned}
P_{\text{succ}}^{\text{PtP-}(u)}=\text{exp}\big(-\dfrac{\tau(R^{(u)}) \sigma^2}{\Omega^{(u)} P}\big),~~u=1,2,
\end{aligned}
\label{eq:P_loss_RLNC_TDMA}
\end{equation}where $\tau(R)=2^R-1$.
 By substituting (\ref{eq:P_loss_RLNC_TDMA}) and \eqref{eq:Nu} into (\ref{eq:ET_RLNC}), the expected completion time in a two-user RLNC-based TDMA scheme is formulated as
\begin{equation}
\begin{aligned}
\mathbb{E}\{T^{\text{TDMA}}\}=\text{exp}&(\dfrac{\tau (R^{(1)}) \sigma^2}{\Omega^{(1)} P}) F(\big\lceil \dfrac{\bar{K}}{R^{(1)}}\big\rceil)\\ &+\text{exp}(\dfrac{\tau(R^{(2)}) \sigma^2}{\Omega^{(2)} P})F(\big\lceil \dfrac{\bar{K}}{R^{(2)}}\big\rceil),
\end{aligned}
\label{eq:ET_RLNC_TDMA}
\end{equation}which is the sum of the expected completion times of PtP transmissions to each user. This is because, in TDMA, only one user is served per time slot. Specifically, the transmitter sends $q_t^{(1)}$ in each slot $t$ until receiving an ACK from user~1, and then serves user~2 by sending $q_t^{(2)}$. Since the transmitter has no knowledge of channel conditions, the same transmission rate is used for both users, i.e., $R^{(1)} = R^{(2)} = R$.
Therefore, the expected completion time (\ref{eq:ET_RLNC_TDMA}) is reduced to 
\begin{equation}
\begin{aligned}
\mathbb{E}&\{T^{\text{TDMA}}\}=\Big(\text{exp}\big(\dfrac{\tau (R) \sigma^2}{\Omega^{(1)}P }\big)+\text{exp}(\dfrac{\tau (R) \sigma^2}{\Omega^{(2)}P })\Big)F(\big\lceil \dfrac{\bar{K}}{R}\big\rceil),
\end{aligned}
\label{eq:ET_RLNC_TDMA2}
\end{equation} implying that the completion time increases exponentially  as the channel gain drops for at least one of the users.

\begin{figure*}
    \centering
    \includegraphics[width=0.85\linewidth]{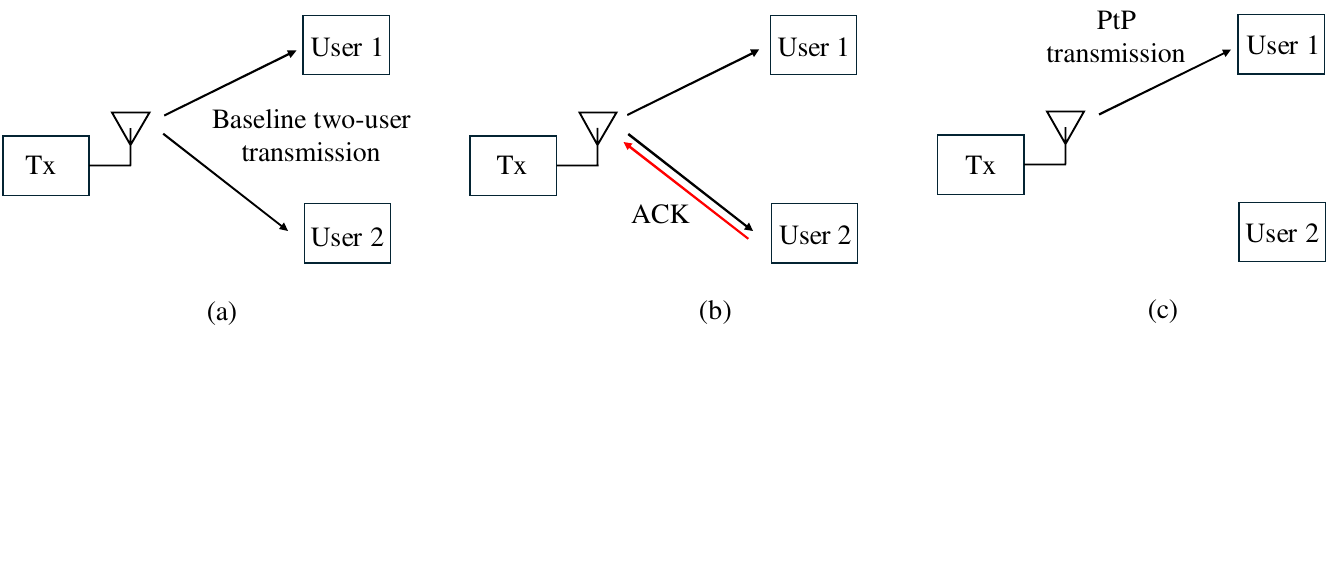}
    \vspace{-3.2cm}
    \caption{Baseline transmission scheme with successive refinement with user~1 weaker than  user~2:
(a) Phase~1: baseline two-user transmission intended for both users;
(b) feedback of an ACK to the transmitter once the decoding matrix at user~2 becomes full rank; and
(c) Phase~2: PtP transmission to user~1.}
    \vspace{-0.2cm}
    \label{fig:baseline}
\end{figure*}

\subsection{RLNC-based  Multicasting}
%..... calculate the latency

In RLNC-based multicasting scheme, the transmitter jointly network-encodes the data streams of both users into a common packet, which is decoded in its entirety by each user. Let the transmission rates of the first and second users be $R^{(1)} = \alpha R$ and $R^{(2)} = (1-\alpha)R$, respectively, where $\alpha \in [0,1]$ denotes the rate-splitting parameter. Based on the assigned rate $R^{(u)}$, the data stream of user $u$ is segmented into a set of packets $p_1^{(u)}, p_2^{(u)}, \ldots, p_{N^{(u)}}^{(u)}$, where $N^{(u)}=\lceil \bar K/{R^{(u)}}\rceil$ for $u = 1, 2$.
Following the RLNC-based encoding procedure in \eqref{eq:RLNC-packet-
TDMA}, the network layer independently generates the encoded packets $q_t^{(1)}$ and $q_t^{(2)}$ at each time slot $t$, together with their corresponding coding vectors $a_{t1}^{(u)}, a_{t2}^{(u)}, \ldots, a_{tN^{(u)}}^{(u)}$. These encoded packets are then concatenated to form a unified packet $q_t$, which is subsequently processed at the physical layer and transmitted using the aggregate transmission rate $R$.

At the user side, each user decodes the received packet $q_t$ at the data rate $R$ to recover the entire packet. Upon successful channel-decoding by user $u$, the packet is partitioned into its constituent components $q_t^{(1)}$ and $q_t^{(2)}$ and appends $q_t^{(u)}$ to its decoding matrix. Under a joint-rank decoding policy, the transmission continues until the decoding matrices of both users reach full rank, at which point they each return an ACK to the transmitter.
 Accordingly, the expected completion time in a two-user RLNC-based joint-rank multicast (JR-MC) scheme is written as
\begin{equation}
\begin{aligned}
\mathbb{E}[T^{\text{JR-MC}}]=\max\Big\{&\text{exp}(\dfrac{\tau(R) \sigma^2}{\Omega^{(1)} P}) F(\Big\lceil \dfrac{\bar{K}}{\alpha R}\Big\rceil )\\ &,\text{exp}(\dfrac{\tau(R) \sigma^2}{\Omega^{(2)} P}) F(\Big\lceil \dfrac{\bar{K}}{(1-\alpha) R}\Big\rceil)\Big\},
\end{aligned}
\label{eq:ET_multicast}
\end{equation}implying that, while the aggregate per-packet rate $R$ remains fixed, rate splitting increases the number of packets each user must successfully receive by factors of $\alpha$ and $1-\alpha$ compared to the PtP case.
 Since the transmitter has no channel knowledge, it adopts equal rate splitting with $\alpha = 0.5$. %Then,  (\ref{eq:ET_multicast}) is rewritten as 
%\begin{equation}
%\begin{aligned}
%\mathbb{E}&[T^{\text{JR-MC}}]= \text{exp}\Big( \dfrac{\tau (R) \sigma^2}{\min\{\Omega^{(1)},\Omega^{(2)}\}P} \Big)  F(\lceil \dfrac{2\bar{K}}{R}\rceil),
%\end{aligned}
%\label{eq:ET_multicast2}
%\end{equation}which is dominated by the weaker channel.

Alternatively, the transmitter may employ the successive refinement multicast (SR-MC) scheme to reduce completion time. As shown in Fig.~\ref{fig:baseline}, multicast packets for both users are transmitted in Phase~1, during which user~2 decodes earlier due to its stronger channel and returns an ACK upon reaching full rank, triggering the transition to Phase~2. In Phase~2, the transmitter applies RLNC in \eqref{eq:RLNC-packet- TDMA} to the segmented packets of user~1,
$p_1^{(1)}, \ldots, p_{N^{(1)}}^{(1)}$, generating coded packets $q_t = q_t^{(1)}$, which are channel-encoded at rate $R/2$ and transmitted over a PtP link for decoding by user~1.

 The completion time of Phase~1 in the SR-MC scheme, denoted by $T^{\text{SR-MC}}_{\text{Ph-1}}$, corresponds to the  time at which the decoding matrix of user~2 achieves full rank
 $N^{(2)}=\lceil \bar K/{R^{(2)}}\rceil$, is derived as
\begin{equation}
\begin{aligned}
\mathbb{E}&[T^{\text{SR-MC}}_{\text{Ph-1}}]= \text{exp}\Big( \dfrac{\tau (R) \sigma^2}{\Omega^{(2)}P} \Big)  F\!\Big( \big\lceil \dfrac{2\bar{K}}{R} \big\rceil \Big).
\end{aligned}
\label{eq:ET_SR-multicast-Ph1}
\end{equation} During this interval, however, user~1 typically receives and successfully decodes only a subset of its intended packets, which are then appended to its decoding matrix.

Although user~1’s decoding matrix remains rank-deficient at the end of Phase~1, it significantly reduces the duration of Phase~2. Packets successfully received in Phase~2 are appended to the existing matrix from Phase~1, accelerating the attainment of full rank. Therefore, to evaluate the expected completion time of Phase~2 in the SR-MC scheme, it is necessary to first determine the rank of user~1’s decoding matrix at the end of Phase~1.
 This rank is denoted by $N^{\text{SR-MC-(1)}}_{\text{Ph-1}}$, whose expectation is lower bounded as
\begin{equation}
\begin{aligned}
\mathbb{E}&\{N^{\text{SR-MC-(1)}}_\text{Ph-1}\}> \dfrac{3}{4}\exp\!\bigg( 
\dfrac{-\tau(R)\,\sigma^2}{\Omega_{\text{Ph-1}}P}
\bigg)
F\!\Big( \big\lceil \dfrac{2\bar{K}}{R} \big\rceil \Big),
\end{aligned}
\label{eq:EN_RLNC_SR-MC-Phase1}
\end{equation}
where
\begin{equation*}
\begin{aligned}
\dfrac{1}{\Omega_{\text{Ph-1}}}=\dfrac{1}{\Omega^{(1)}} - \dfrac{1}{\Omega^{(2)}},
\end{aligned}
\label{eq:where}
\end{equation*}
whose proof is provided in Appendix \ref{appendix:rank-SR-MC}. The rank of the decoding matrix of user~1 at the beginning of Phase~2 of the SR-MC transmission scheme is identical to its rank at the end of Phase~1. During Phase~2, the transmitter encodes only the packets intended for user~1 and transmits them over a PtP channel, such that the rank of the decoding matrix increases from $N^{\text{SR-MC-(1)}}_{\text{Ph-1}}$ to 
$N^{(1)} = \lceil 2\bar{K}/R\rceil$.

Since the data stream of user~1 is segmented into 
$N^{(1)} = \lceil 2\bar{K}/R\rceil$ packets according to the rate $R/2$ in Phase~1, the same packets are subsequently network-encoded in Phase~2. Then, user~1 is required to decode only its own packets at a data rate $R/2$ during Phase~2.
So, the expected transmission time of Phase 2 of SC-MC scheme is upper-bounded as 
\begin{equation}
\begin{aligned}
\mathbb{E}\!\big\{& T^{\textnormal{SR-MC}}_{\textnormal{Ph-2}}\big\}<\exp\!\Big(\frac{\tau(R/2) \sigma^2}{\Omega^{(1)}  P}\Big)
\\&
\times \bigg(1-\dfrac{3}{4}\exp\!\Big( 
\dfrac{-\tau(R)\,\sigma^2}{\Omega_{\text{Ph-1}}P}
\Big)
\bigg)
F\!\Big( \big\lceil \dfrac{2\bar{K}}{R} \big\rceil \Big),
\end{aligned}
\label{eq:ET_RLNC_SR-MC_Phase2-user1}
\end{equation}
whose proof is provided in Appendix \ref{Appendix:ET-phase2}. Finally, the total expected transmission time of SC-MC scheme is derived as 
\begin{equation}
\begin{aligned}
\mathbb{E}\!\big\{ T^{\textnormal{SR-MC}}\big\}=\mathbb{E}\!\big\{ T^{\textnormal{SR-MC}}_{\textnormal{Ph-1}}\big\}+ \mathbb{E}\!\big\{& T^{\textnormal{SR-MC}}_{\textnormal{Ph-2}}\big\},
\end{aligned}
\label{eq:ET_RLNC_SR-MC_total}
\end{equation}which is upperbounded by the sum of the right hand terms in (\ref{eq:ET_SR-multicast-Ph1}) and (\ref{eq:ET_RLNC_SR-MC_Phase2-user1})

\subsection{RLNC-based FDMA}
In the RLNC-based FDMA scheme, the available bandwidth is equally shared among users, while the packet size and total system bandwidth remain identical to the PtP case. Consequently, in each FDMA time slot, each user is allocated half of the packet symbols, i.e., $n_{\text{FFT}}/2$.
At the network layer, each user’s data stream is segmented at rate $R^{(u)}$ into $N^{(u)} = \lceil \bar{K}/R^{(u)} \rceil$ packets, and RLNC-encoded packets $q_t^{(u)}$ are generated independently using \eqref{eq:RLNC-packet- TDMA}. Each $q_t^{(u)}$ is encoded and modulated at rate $R^{(u)}$ to form packets of length $n_{\text{FFT}}$, which are then split into two halves of length $n_{\text{FFT}}/2$. Two composite packets are constructed by concatenating the corresponding halves of users $u=1$ and $u=2$ and transmitted over two consecutive time slots.

Given a total transmit power budget of $P$, equally dividing the bandwidth allocates $P/2$ power to each subband. Since thermal noise power scales with bandwidth, the noise power at each receiver is also reduced by half. Therefore, the success probability of a half-packet, denoted as $P_{\text{succ, half}}^{\text{FDMA-}(u)}$, follows the PtP packet success probability in \eqref{eq:P_loss_RLNC_TDMA}. Consequently, the decoding matrix rank at each user increases only after successful reception of a complete packet formed by two halves transmitted in consecutive time slots.
Therefore, the per-packet success probability is written as 
\begin{equation}
\begin{aligned}
      P_{\text{succ}}^{\text{FDMA-}(u)}&=  (P_{\text{succ, half}}^{\text{FDMA-}(u)})^2
       \\& =\text{exp}(\dfrac{-2\,\tau(R^{(u)}) \sigma^2}{\Omega^{(u)} P})~~~~~\text{for}~u=1,2.
\end{aligned}\label{eq:full_FDMA}
\end{equation}

By inserting the per-packet success probability \eqref{eq:full_FDMA} into (\ref{eq:ET_RLNC}), we derive the  expected completion time of a two-user joint rank FDMA (JR-FDMA) scheme as
\begin{equation}
\begin{aligned}
\mathbb{E}\{T^{\text{JR-FDMA}}\}=2\max\Big\{&\text{exp}(\dfrac{2\,\tau(R^{(1)}) \sigma^2}{\Omega^{(1)} P})F(\big\lceil \dfrac{\bar{K}}{R^{(1)}}\big\rceil)\\ &,\text{exp}(\dfrac{2\,\tau(R^{(2)}) \sigma^2}{\Omega^{(2)} P})F(\big\lceil \dfrac{\bar{K}}{R^{(2)}}\big\rceil)\Big\},
\end{aligned}
\label{eq:ET_RLNC_FDMA}
\end{equation}where the factor of two reflects that each complete packet requires two time slots per user. Also, due to no channel knowledge, the data rates are set as
, i.e., $R^{(1)} = R^{(2)} = R$.
% Therefore,  (\ref{eq:ET_RLNC_FDMA}) is reduced to
%\begin{equation}
%\begin{aligned}
%\mathbb{E}& \{T^{\text{JR-FDMA}}\}= 2 ~\text{exp}\Big( \dfrac{2\,\tau (R) \sigma^2}{\min\{\Omega^{(1)},\Omega^{(2)}\}P}  \Big)F\big(\big\lceil \dfrac{\bar{K}}{R}\big\rceil\big).
%\end{aligned}
%\label{eq:ET_RLNC_FDMA2}
%\end{equation}

To further reduce completion time, the transmitter may employ the successive refinement FDMA (SR-FDMA) scheme. Similar to the arguments on Phase 1 of SR-MC scheme, % As shown in Fig.~\ref{fig:baseline}a), Phase~1 uses FDMA to transmit packets to both users and terminates upon receiving an ACK from user~2 when its decoding matrix reaches full rank  $N^{(2)}=\lceil \bar K/{R^{(2)}}\rceil=\lceil \bar K/{R}\rceil$. 
the expected completion time in Phase~1 of the SR-FDMA scheme, denoted by $T^{\text{SR-FDMA}}_{\text{Ph-1}}$,  is calculated as 
\begin{equation}
\begin{aligned}
\mathbb{E}&[T^{\text{SR-FDMA}}_{\text{Ph-1}}]= 2\,\text{exp}\Big( \dfrac{2\,\tau (R) \sigma^2}{\Omega^{(2)}P} \Big)  F\!\Big( \big\lceil \dfrac{\bar{K}}{R} \big\rceil \Big).
\end{aligned}
\label{eq:ET_SR-FDMA-Ph1}
\end{equation}
To determine the expected completion time of Phase~2 of the SR-FDMA scheme, we first  characterize the rank of the decoding matrix of user~1 at the end of Phase~1. Denoting this rank by $N^{\text{SR-FDMA-(1)}}_{\text{Ph-1}}$ and using arguments analogous to those presented in Appendix~A, its expectation is lower bounded as
\begin{equation}
\begin{aligned}
\mathbb{E}&\{N^{\text{SR-FDMA-(1)}}_\text{Ph-1}\}> \dfrac{3}{4}\exp\!\bigg( 
\dfrac{-2\,\tau(R)\,\sigma^2}{\Omega_{\text{Ph-1}}P}
\bigg)
F\!\Big( \big\lceil \dfrac{\bar{K}}{R} \big\rceil \Big).
\end{aligned}
\label{eq:EN_RLNC_SR-FDMA-Phase1}
\end{equation}

%During Phase~2, the transmitter encodes only packets intended for user~1 and transmits them over a PtP channel, increasing the rank of user~1’s decoding matrix from $N^{\text{SR-FDMA-(1)}}_{\text{Ph-1}}$ to$N^{(1)} = \lceil \bar{K}/R \rceil$. %Since user~1’s data stream is segmented into $N^{(1)}$ packets in Phase~1, the same packets are RLNC-encoded in Phase~2. 
Thus, using \eqref{eq:EN_RLNC_SR-FDMA-Phase1} and arguments similar to those for Phase~2 of SC-MC and in Appendix~B, the expected transmission time of Phase~2 of the SR-FDMA scheme is upper bounded as
\begin{equation}
\begin{aligned}
\mathbb{E}\!\big\{& T^{\textnormal{SR-FDMA}}_{\textnormal{Ph-2}}\big\}<\exp\!\Big(\frac{\tau(R) \sigma^2}{\Omega^{(1)}  P}\Big)
\\&
\times \bigg(1-\dfrac{3}{4}\exp\!\Big( 
\dfrac{-2\tau(R)\,\sigma^2}{\Omega_{\text{Ph-1}}P}
\Big)
\bigg)
F\!\Big( \big\lceil \dfrac{\bar{K}}{R} \big\rceil \Big).
\end{aligned}
\label{eq:ET_RLNC_SR-FDMA_Phase2-user1}
\end{equation}
 Finally, the expected completion time of a two-user SC-FDMA scheme, denoted by $ T^{\textnormal{SR-FDMA}}$, is derived as the sum of the expected transmission times in each phase, which is upper bounded by the sum of the right-hand sides of (\ref{eq:ET_SR-FDMA-Ph1}) and (\ref{eq:ET_RLNC_SR-FDMA_Phase2-user1}).

\subsection{RLNC-based Inter-Session}
In the RLNC-based inter-session transmission scheme, the transmitter jointly encodes the packets belonging to both users. So, a network-encoded packet at time slot $t$ is generated as
\begin{equation}
\begin{aligned}
q_t=\bigoplus_{u=1}^{2}\bigoplus_{\substack{i=1 }}^{N} a_{ti}^{(u)}p_i^{(u)},
\end{aligned}
\label{eq:IS_RLNC-packet}
\end{equation}
where  $a_{ti}^u=\{0,1\}$ has a symmetric Bernoulli distribution. Since each packet $q_t$ is formed as the XOR of packets from both users, the packet size $p_i^{(u)}$ must be identical for $u=1$ and $u=2$. This implies the same number of packets at the network layer, i.e., $N^{(1)} = N^{(2)}$, given the equal data stream size $K$ for both users. As $N^{(u)} = \lceil \bar{K}/R^{(u)} \rceil$, the transmission rates must be equal, yielding $R^{(1)} = R^{(2)} = R$. Hence, the number of packets per user is
$N^{(1)} = N^{(2)} = N = \lceil \bar{K}/R \rceil$,
and each packet $q_t$ is encoded and modulated at rate $R$ for transmission in time slot $t$.

At user side, the receiving packets are decoded independently at each user with the data rate $R$.  In this setting, the packet loss probability follows the PtP model  characterized by (\ref{eq:PtP_P_loss}). Therefore, the expected completion time in a two-user inter-session RLNC scheme is derived as 
\begin{equation}
\begin{aligned}
\mathbb{E}\{T^{\text{IS}}\}= \text{exp}\Big( \dfrac{\tau (R) \sigma^2}{\min\{\Omega^{(1)},\Omega^{(2)}\}P} \Big)  F(2\lceil \bar{K}/R\rceil),
\end{aligned}
\label{eq:ET_RLNC_IS}
\end{equation}where the factor $2$ accounts for the requirement that each user’s decoding matrix reaches rank $2N$ to enable packet recovery. 
Therefore, the inter-session transmission scheme terminates only after both users return acknowledgement, since successful recovery requires decoding all packets from both users.

\begin{figure*}
    \centering
    \includegraphics[width=0.8\linewidth]{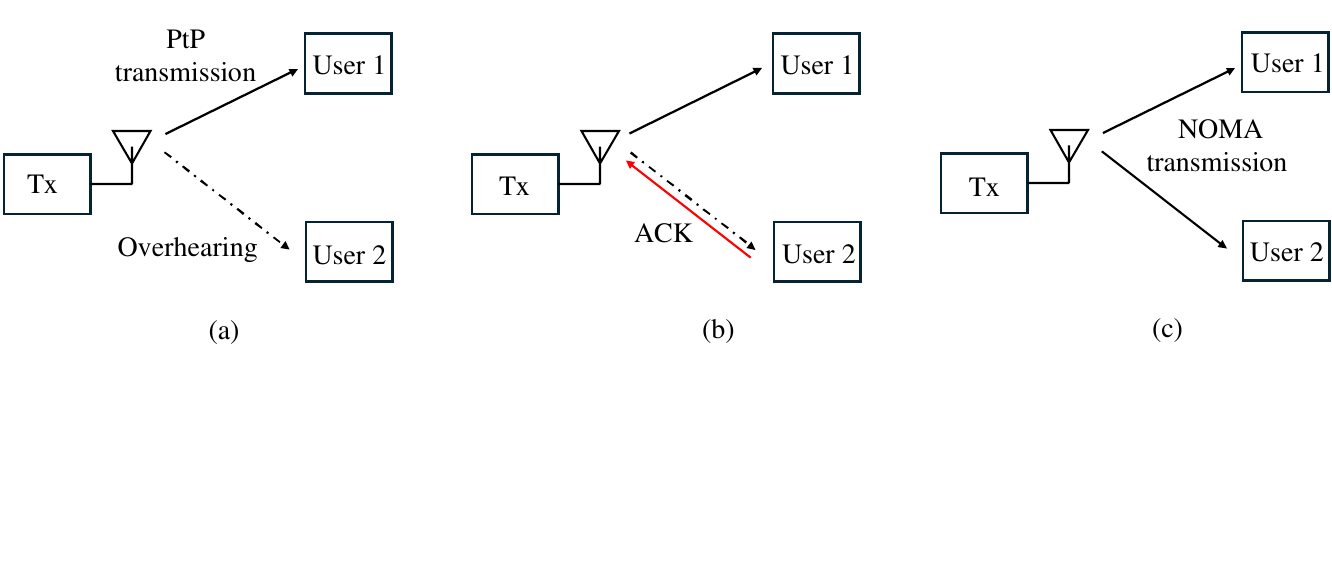}
    \vspace{-2.7cm}
    \caption{ ONOMA transmission scheme with user~1 weaker than user~2:
(a) Phase~1: PtP packet transmission to user~1, while user~2 overhears the transmission (dash-dotted line);
(b) Acknowledgment once the decoding matrix at user~2 becomes full rank; and
(c) Phase~2: symbol-aware NOMA transmission.}
    \vspace{-0.2cm}
    \label{fig:proposed}
\end{figure*}

\subsection{RLNC-based  Classical NOMA}
In the RLNC-based NOMA transmission scheme, the transmitter follows~(\ref{eq:RLNC-packet-
TDMA}) at the network layer to independently encode the packets for each user. Then, at each time slot $t$, the network-encoded packets $q_t^{(1)}$ and $q_t^{(2)}$ are independently channel-encoded and modulated  with data rate $R^{(u)}$, resulting in the symbol sequences
$x_{t1}^{(u)}, x_{t2}^{(u)}, \ldots, x_{t n_{\mathrm{FFT}}}^{(u)}$ for $u=1,2$. 
The transmitted signal is then constructed by superimposing the symbols corresponding to both users.
 Then, the received symbol at each user is described as
\begin{equation}
    y^{(u)}=h^{(u)}(\sqrt{\lambda}x^{(1)}+\sqrt{1-\lambda}x^{(2)})+n^{(u)}, ~\text{for}~u=1,2
    \label{eq:NOMA-symbols}
\end{equation}where $x^{(u)}:=x_{ti}^{(u)}$ for $i=1,2,\ldots, n_{\text{FFT}}$, denotes the symbol of user $u$ with $\mathbb{E}\{|x^{{(u)}}|^2\}=P$ for $u=1,2$. 

At the user side, user~1 decodes its own message by treating the signal of the second user as noise. Thus, the probability of successful packet decoding at user~1 is given by
\begin{equation}
\begin{aligned}
P_{\text{succ}}^{\text{NOMA-(1)}}
&= \Pr\!\left(\frac{\lambda |h^{(1)}|^2 P}{(1-\lambda)|h^{(1)}|^2 P + \sigma^2} >\tau(R^{(1)}) \right)
 \\
&= \exp\!\left(-\frac{\tau(R^{(1)}) \sigma^2}{\Omega^{(1)} \big(\lambda - \tau(R^{(1)}) (1-\lambda)\big)P}\right).
\end{aligned}
\label{eq:P_loss_NOMA1}
\end{equation}
 In contrast, user 2 employs successive interference cancellation (SIC) to first decode the signal of  user 1, subtract it from the received signal, and subsequently decode its own message. Hence, channel decoding at user~2 is successful if both decoding phases succeed, whose probability is written as 
\begin{equation}
\begin{aligned}
P_{\text{succ}}^{\text{NOMA-(2)}} =
 \Pr\Big(&\dfrac{\lambda |h^{(2)}|^2 P}{(1-\lambda)|h^{(2)}|^2 P + \sigma^2} >\tau(R^{(1)}),  \\
&, \dfrac{(1-\lambda) |h^{(2)}|^2 P}{\sigma^2} >\tau(R^{(2)})\Big)
 \\
&\hspace{-2.6cm}= \exp\!\Big(\dfrac{-\sigma^2}{\Omega^{(2)}P} \times \max\Big\{\dfrac{\tau(R^{(1)})}{ \lambda -\tau(R^{(1)}) (1-\lambda)} ,\dfrac{\tau(R^{(2)}) }{ (1-\lambda) }\Big\}\Big),
\end{aligned}
\label{eq:P_loss_NOMA2}
\end{equation}
 where  $\lambda$ must satisfy
\eqref{eq:lambda-requirement}.
Since the transmitter has no channel knowledge, it employs identical data rates $R^{(1)} = R^{(2)} = R < 1$ and sets $\lambda = 0.5$, under which the left-hand term in~(\ref{eq:P_loss_NOMA2}) dominates. Substituting~(\ref{eq:P_loss_NOMA1}) and~(\ref{eq:P_loss_NOMA2}) into~(\ref{eq:ET_RLNC}), the expected completion time of the two-user RLNC-based NOMA scheme with a joint-rank policy is obtained as
\begin{equation}
\begin{aligned}
&\mathbb{E}\{T^{\text{JR-NOMA}}\}\\&=\exp \Big(\frac{2\tau(R) \sigma^2}{\min\{\Omega^{(1)},\Omega^{(2)}\}\big(1-\tau(R)\big)P}\Big)F\big(\lceil \dfrac{\bar{K}}{R}\rceil\big),
\end{aligned}
\label{eq:ET-JR-NOMA}
\end{equation} where $\tau(R)<1$. So, the data rate must be set as $R<1$. %Under the joint-rank policy, transmission terminates only after both users have returned an ACK.

To reduce the overall completion time, the transmitter may employ the successive-refinement NOMA (SR-NOMA). In Phase~1, assuming that $\Omega^{(1)}<\Omega^{(2)}$, the packets intended for both users are transmitted using the NOMA scheme until user~2 returns an ACK. %This ACK indicates that its decoding matrix achieves full rank $N^{(2)} = \lceil \bar{K}/{R^{(2)}} \rceil = \lceil \bar{K}/{R} \rceil$, allowing the transmitter to switch to Phase~2.
 %The transmitter then switches to Phase~2, where only user~1’s packets are transmitted. In Phase~2, the transmitter exclusively encodes the packets $p_1^{(1)}, \ldots, p_{N^{(1)}}^{(1)}$, with $N^{(1)} = \lceil \bar{K}/R \rceil$, using the RLNC procedure in~\eqref{eq:RLNC-packet- TDMA}, and delivers them over a PtP channel at data rate $R$ for decoding solely by user~1.
%So, the completion time of Phase~1 in the SR-NOMA scheme, denoted by$T^{\text{SR-NOMA}}_{\text{Ph-1}}$, is defined as the time slot at which the decoding matrix of user~2 attains full rank $N^{(2)} = \lceil \bar{K}/{R^{(2)}} \rceil= \lceil \bar{K}/{R} \rceil .$
Therefore, by invoking~\eqref{eq:ET_RLNC} and \eqref{eq:ET-JR-NOMA}, the expected transmission time of Phase~1, denote by $T^{\text{SR-NOMA}}_{\text{Ph-1}}$, is obtained as
\begin{equation}
\begin{aligned}
\mathbb{E}[T^{\text{SR-NOMA}}_{\text{Ph-1}}]= \exp \Big(\frac{2\tau(R) \sigma^2}{\Omega^{(2)}\big(1-\tau(R)\big)P}\Big)F\big(\lceil \bar{K}/R\rceil\big).
\end{aligned}
\label{eq:ET_SR-NOMA-Ph1}
\end{equation}
%During this interval, user~1 may receive and successfully decode a subset of its intended packets, which are subsequently incorporated into its decoding matrix.
Denoting the rank of the decoding matrix at user~1 at the end of Phase~1 by
$N^{\text{SR-NOMA-(1)}}_{\text{Ph-1}}$, its expected value can be lower bounded as
\begin{equation} \begin{aligned} \mathbb{E}&\{N^{\text{SR-NOMA-(1)}}_\text{Ph-1}\}> \dfrac{3}{4}\exp \Big(\frac{-2\tau(R) \sigma^2}{\Omega_{\text{Ph-1}}\big(1-\tau(R)\big)P}\Big) F\!\Big( \big\lceil \dfrac{\bar{K}}{R} \big\rceil \Big), \end{aligned} \label{eq:EN_RLNC_SR-NOMA-Phase1} 
\end{equation}
which follows from \eqref{eq:P_loss_NOMA1} and
\eqref{eq:ET_SR-NOMA-Ph1} by setting $\lambda = 0.5$, and by using arguments
similar to those presented in Appendix~\ref{appendix:rank-SR-MC}.

During Phase~2, the transmitter encodes only packets intended for user~1 and transmits them over a PtP channel, increasing the rank of user~1’s decoding matrix from
$N^{\text{SR-NOMA-(1)}}_{\text{Ph-1}}$ to
$N^{(1)} = \lceil \bar{K}/R \rceil$.
%Since user~1’s data stream is segmented into $N^{(1)}$ packets in Phase~1, the same packets are network-encoded in Phase~2.
So, using \eqref{eq:EN_RLNC_SR-NOMA-Phase1} and arguments analogous to Appendix~B, the expected transmission time of Phase~2 of the SR-NOMA scheme is upper bounded as
\begin{equation}
\begin{aligned}
\mathbb{E}\!\big\{& T^{\textnormal{SR-NOMA}}_{\textnormal{Ph-2}}\big\}<\exp\!\Big(\frac{\tau(R) \sigma^2}{\Omega^{(1)}  P}\Big)
\\&
\times \bigg(1- \dfrac{3}{4}\exp \Big(\frac{-2\tau(R) \sigma^2}{\Omega_{\text{Ph-1}}\big(1-\tau(R)\big)P}\Big)
\bigg)
F\!\Big( \big\lceil \dfrac{\bar{K}}{R} \big\rceil \Big).
\end{aligned}
\label{eq:ET_RLNC_SR-NOMA_Phase2-user1}
\end{equation}
Finally, the expected completion time of the two-user SC-NOMA scheme, denoted by $T^{\textnormal{SR-NOMA}}$, is obtained as the sum of the expected transmission times in each phase and is upper bounded by the sum of the right-hand sides of (\ref{eq:ET_SR-NOMA-Ph1}) and (\ref{eq:ET_RLNC_SR-NOMA_Phase2-user1}).
 %at each phase as
%\begin{equation}
%\begin{aligned}
%\mathbb{E}\!\big\{ T^{\textnormal{SR-NOMA}}\big\}=\mathbb{E}\!\big\{ T^{\textnormal{SR-NOMA}}_{\textnormal{Ph-1}}\big\}+ \mathbb{E}\!\big\{& T^{\textnormal{SR-NOMA}}_{\textnormal{Ph-2}}\big\},
%\end{aligned}
%\label{eq:ET_RLNC_SR-FDMA_total}
%\end{equation}which is upperbounded by the sum of the right hand terms in (\ref{eq:ET_SR-NOMA-Ph1}) and (\ref{eq:ET_RLNC_SR-NOMA_Phase2-user1}).

%\subsection{RSMA}
%..... calculate the latency 

\section{Proposed transmission protocol: RLNC-based Overhearing-driven NOMA (ONOMA)}\label{sec:ONOMA}

In the proposed RLNC-based overhearing-driven NOMA (ONOMA) scheme, the transmitter leverages overhearing and ACK timing to infer channel conditions and adjust transmission parameters. Unlike classical NOMA, where the strong user performs decoding-based SIC within the same time slot, ONOMA relies on prior side information from an explicit overhearing phase. The stronger user therefore applies reconstruction-based interference subtraction instead of conventional SIC, so performance gains arise from overhearing-induced information asymmetry rather than power-domain multiplexing alone. As in the baseline setting, we consider a two-user network in which each user has $K$ bits to transmit, with $\Omega^{(1)} < \Omega^{(2)}$, while the transmitter has no prior knowledge of the channel conditions. In the following, each phase of the ONOMA transmission scheme is described in detail.

%The joint-rank ONOMA (JR-ONOMA) scheme, shown in Fig.~\ref{fig:proposed}, operates in two consecutive phases. In Phase~1, the transmitter sends user~1’s packets over a PtP channel while user~2 passively overhears and decodes them. When the decoding matrix at user~2 reaches full rank, it sends an ACK, having recovered all packets of user~1. Phase~2 then begins, where the packets of both users are transmitted using NOMA. In the following, each phase is described in detail.

\subsection{Phase 1: PtP transmission and overhearing}
In this phase, as illustrated in Fig.~\ref{fig:proposed}a), the transmitter partitions the data streams intended for each user into
$N^{(1)} = N^{(2)} = N = \big\lceil \bar{K}/R \big\rceil$ packets and applies network coding exclusively to the packets of user~1, i.e.,
$p_1^{(1)}, p_2^{(1)}, \ldots, p_{N}^{(1)}$, via~\eqref{eq:RLNC-packet-
TDMA}. 
Each packet is subsequently channel encoded at the physical layer with data rate~$R$ and transmitted over a PtP channel for decoding at user~1, while user~2 overhears and attempts to decode the same packets.

At the user side, each user independently constructs a decoding matrix from the received packets and appends a packet to the matrix whenever it is successfully decoded. During this phase, the packet success probability for each user, denoted by $P_{\text{succ,Ph-1}}^{\text{ONOMA-(u)}}$,  follows PtP packet success probability  presented in \eqref{eq:P_loss_RLNC_TDMA}. %and is given by
%\begin{equation}
%P_{\text{succ,Ph-1}}^{\text{ONOMA-(u)}}
%=\exp\!\left(\frac{-\tau(R)\sigma^2}{\Omega^{(u)}P}\right), 
%\quad u=1,2 .
%\label{eq:P_loss_Ph1}
%\end{equation}
However, the decoding matrix at user~2 is expected to reach full rank $N$ earlier than that of user~1.
As soon as achieving the full rank, an ACK is sent to the transmitter, thereby terminating Phase~1 of the transmission, as illustrated in Fig.~\ref{fig:proposed}b).
Therefore, at the end of Phase~1, user~2 has successfully decoded all packets of user~1, namely,
$p_1^{(1)}, p_2^{(1)}, \ldots, p_N^{(1)}$.
In contrast, the rank of the decoding matrix at user~1 has reached a value
$N_{\text{Ph-1}}^{\text{ONOMA-(1)}} < N$.

\begin{figure*}
    \centering
    \vspace{-2cm}
    \includegraphics[width=0.8\linewidth]{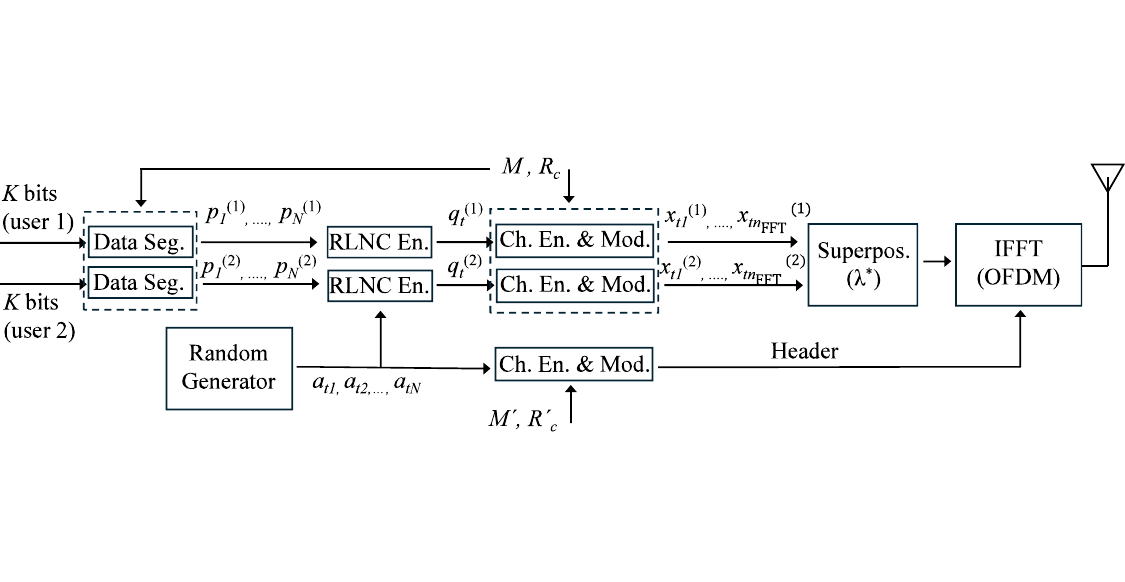}
    \vspace{-2.1cm}
    \caption{Building blocks of the transmitter in Phase 2 of a two-user RLNC-based ONOMA transmission scheme}
    \vspace{-0.5cm}
    \label{fig:phase2}
\end{figure*}

\subsection{Phase 2: NOMA transmission with symbol awareness}

At the start of Phase~2, as shown in Fig.~\ref{fig:proposed}c), each user’s packets are independently network-encoded to produce $q_t^{(1)}$ and $q_t^{(2)}$ using the RLNC procedure in~\eqref{eq:RLNC-packet- TDMA}.
To reduce header overhead, the same random coding vector is used for both users at each time slot, i.e., $a_{ti}^{(1)} = a_{ti}^{(2)} = a_{ti}$ for all $i=1,\ldots,N$. The encoded packets, along with their coding vectors, are then passed to the physical layer for transmission.

Fig.~\ref{fig:phase2} shows the Phase~2 transmitter. Each user’s packets are channel-encoded at rate~$R$, producing symbols $x_{t1}^{(u)}, \ldots, x_{tn_{\text{FFT}}}^{(u)}$, denoted by $x^{(u)}$, for $u=1,2$. The transmitted signal is formed by superimposing $x^{(1)}$ and $x^{(2)}$ with a power allocation factor $\lambda > 0.5$, since user~1 is now identified as the weaker user.
 The coding vector, smaller than the payload, is encoded at a lower rate~$R'$ and appended before the IFFT, with $R' < R$, to ensure a reliable header delivery.

At each user, the received signal is the superposition of both users’ symbols, as given in~\eqref{eq:NOMA-symbols}. User~2 first decodes the header to obtain the coding vector. Since the packets of user~1, $p_1^{(1)}, \ldots, p_N^{(1)}$, are already known to user~2, it applies the same RLNC procedure as the transmitter to reconstruct $q_t^{(1)}$ according to~\eqref{eq:RLNC-packet- TDMA}.
Given the known modulation order~$M$ and code rate~$R_c$, user~2 regenerates the transmitted symbols $x_{t1}^{(1)}, \ldots, x_{tn_{\text{FFT}}}^{(1)}$ and cancels it from its received signal via SIC\footnote{Security is preserved under standard layered encryption assumptions.}. Thus, the received signal expressions in~\eqref{eq:NOMA-symbols} are rewritten as
\begin{align}
    y^{(1)} &= h^{(1)}\!\left(\sqrt{\lambda}\,x^{(1)} + \sqrt{1-\lambda}\,x^{(2)}\right) + n^{(1)},
    \label{eq:Phase2-symbols-user1}
\end{align}
and
\begin{align}
    y^{(2)} &= h^{(2)} \sqrt{1-\lambda}\,x^{(2)} + n^{(2)},
    \label{eq:Phase2-symbols-user2}
\end{align} respectively. Therefore, user 2 is capable of decoding its own packets with no interference. Accordingly, the probability of successful packet decoding at user 1 and 2 are formulated as
\begin{equation}
\begin{aligned}
P_{\text{succ, Ph-2}}^{\text{ONOMA-(1)}}
&= \Pr\!\left(\frac{\lambda |h^{(1)}|^2 P}{(1-\lambda)|h^{(1)}|^2 P + \sigma^2} >\tau(R) \right)
 \\
&= \exp\!\left(-\frac{\tau(R) \sigma^2}{\Omega^{(1)} (\lambda - \tau(R) (1-\lambda))P}\right).
\end{aligned}
\label{eq:P_loss_Phase2-1-user-1}
\end{equation}
and 
\begin{equation}
\begin{aligned}
P_{\text{succ, Ph-2}}^{\text{ONOMA-(2)}}
&= \Pr\!\left(\frac{(1-\lambda) |h^{(2)}|^2 P}{  \sigma^2} >\tau(R) \right)
 \\
&= \exp\!\left(-\frac{\tau(R) \sigma^2}{(1-\lambda)\Omega^{(2)}P}\right),
\end{aligned}
\label{eq:P_loss_Phase2-1-user-2}
\end{equation}respectively. From~\eqref{eq:P_loss_Phase2-1-user-1} and~\eqref{eq:P_loss_Phase2-1-user-2}, a too small $\lambda$ severely degrades the packet success probability of user~1, whereas $\lambda$ close to one significantly reduces that of user~2. So, $\lambda$ must be carefully selected to ensure efficient ONOMA transmission. Both users are expected to successfully decode their packets during Phase~2 with the proposed symbol-aware NOMA transmission, even if their channel conditions switch. This is because user~1, when experiencing favorable channel conditions, receives a larger portion of the transmit power, while user~2, even with a weak channel, performs interference-free decoding after symbol reconstruction.

At the start of Phase~2, user~1 has decoding matrix rank $N_{\text{Ph-1}}^{\text{ONOMA-(1)}}$,  built in Phase~1. Each successfully decoded packet in Phase~2 adds its coding vector to the matrix, and user~1 sends an ACK once the rank reaches~$N$.
In contrast, user~2 begins Phase~2 with an empty matrix and appends a coding vector whenever it successfully decodes a packet.

In the JR-ONOMA scheme, Phase~2 continues until ACKs are received from both users, ensuring that user~1, with the weaker channel, receives its packets throughout both phases.
In contrast, in the SR-ONOMA scheme, Phase~2 ends once either user attains full rank and sends an ACK. Phase~3 then begins, where the rank-deficient user’s packets are transmitted over a PtP channel until acknowledgment. %Due to this early termination of Phase~2, SR-ONOMA is expected to achieve a smaller completion time than JR-ONOMA.

\subsection{Latency Analysis}

 Since $\Omega^{(1)} < \Omega^{(2)}$, user~2 returns ACK earlier, and thus, the completion time $T_{\mathrm{Ph\text{-}1}}^{\mathrm{ONOMA}}$ is determined by the time needed for user~2’s decoding matrix to reach rank $N$ in Phase~1. Its expected value is given by
\begin{equation}
\begin{aligned}
\mathbb{E}&\{T_\text{Ph-1}^{\text{ONOMA}}\}=\text{exp}\big( \dfrac{\tau (R) \sigma^2}{\Omega^{(2)}P}\big)F(\big\lceil \dfrac{\bar{K}}{R}\big\rceil),
\end{aligned}
\label{eq:ET_RLNC_Phase1}
\end{equation}

By the end of Phase~1, the rank of the decoding matrix at user~1 corresponding to the decoding of its own packets reaches a value denoted by $N_{\text{Ph-1}}^{\text{ONOMA-(1)}}<N$.
Accordingly, using \eqref{eq:ET_RLNC_Phase1} and   similar arguments presented in Appendix~\ref{appendix:rank-SR-MC}, the expected value of this rank can be  lower bounded as
\begin{equation}
\begin{aligned}
\mathbb{E}\!\big\{& N^{\textnormal{ONOMA-(1)}}_{\textnormal{Ph-1}}\big\}
 >\dfrac{3}{4}
\exp\!\bigg( 
\dfrac{-\tau(R)\,\sigma^2}{\Omega_{\text{Ph-1}}P}
\bigg)
F\!\Big( \big\lceil \dfrac{\bar{K}}{R} \big\rceil \Big).
\end{aligned}
\label{eq:EN_RLNC_Phase1-expanded}
\end{equation}
At the start of Phase~2, the transmitter superimposes both users’ symbols. Each user decodes its own symbols according to \eqref{eq:Phase2-symbols-user1} and \eqref{eq:Phase2-symbols-user2}, appending the decoding vector to its matrix upon successful reception.
Let $T_{\text{Ph-2}}^{\text{ONOMA-(1)}}$ denote the time required for user~1’s decoding matrix rank to increase from $N_{\text{Ph-1}}^{\text{ONOMA-(1)}}$ to $N$. Using \eqref{eq:P_loss_Phase2-1-user-1} and \eqref{eq:EN_RLNC_Phase1-expanded}, and arguments similar to Appendix~B, its expected value can be  upper bounded as
\begin{equation}
\begin{aligned}
\mathbb{E}\!\big\{ T^{\textnormal{ONOMA-(1)}}_{\textnormal{Ph-2}}\big\}&<\exp\!\Big(\frac{\tau(R) \sigma^2}{\Omega^{(1)} \lambda' P}\Big)
\\&
\times \bigg(1-\dfrac{3}{4}\exp\!\Big( 
\dfrac{-\tau(R)\,\sigma^2}{\Omega_{\text{Ph-1}}P}
\Big)
\bigg)
F\!\Big( \big\lceil \dfrac{\bar{K}}{R} \big\rceil \Big),
\end{aligned}
\label{eq:ET_RLNC_Phase2-user1}
\end{equation}
where 
\begin{equation*}
\begin{aligned}
\lambda'=\lambda - \tau(R) (1-\lambda).
\end{aligned}
\label{eq:}
\end{equation*}

At the start of Phase~2, user~2’s decoding matrix has rank zero, as no packets for user~2 were sent in Phase~1. During Phase~2, user~2 receives NOMA packets, cancels user~1’s symbols, and decodes its own packet. Upon successful decoding, the corresponding vector is appended to its decoding matrix. User~2 sends an ACK once the rank reaches $N$.
Let $T^{\textnormal{ONOMA-(2)}}_{\textnormal{Ph-2}}$ denote the time required for the decoding matrix at user~2 to grow from rank zero to $N$. Using (\ref{eq:ET_RLNC}), its expected value is given by
\begin{equation}
\begin{aligned}
\mathbb{E}\!\big\{ T^{\textnormal{ONOMA-(2)}}_{\textnormal{Ph-2}}\big\}&=\dfrac{1}{P_{\text{succ, Ph-2}}^{\text{ONOMA-(2)}}}\sum_{r=0}^{N-1}\dfrac{1}{1-2^{r-N}}\\ &=\exp\!\left(\frac{\tau(R) \sigma^2}{(1-\lambda)\Omega^{(2)}P}\right) F\!\Big( \big\lceil \dfrac{\bar{K}}{R} \big\rceil \Big),
\end{aligned}
\label{eq:ET_RLNC_Phase2-user2}
\end{equation}wherein packet success probability in (\ref{eq:P_loss_Phase2-1-user-2}) is used. 
Accordingly, the  expected completion time of the joint rank ONOMA (JR-ONOMA) transmission scheme is given by
\begin{equation}
\begin{aligned}
\mathbb{E}\!\big\{ T^{\textnormal{JR-ONOMA}}\big\}&=  \mathbb{E}\!\big\{ T^{\textnormal{ONOMA}}_{\textnormal{Ph-1}}\big\} + \\& \max \Big\{ \mathbb{E}\!\big\{ T^{\textnormal{ONOMA-(1)}}_{\textnormal{Ph-2}}\big\}, 
\mathbb{E}\!\big\{ T^{\textnormal{ONOMA-(2)}}_{\textnormal{Ph-2}}\big\}\Big\},
\end{aligned}
\label{eq:ET_RLNC_ONOMA-overal}
\end{equation}
whose upperbound is determined by substituting (\ref{eq:ET_RLNC_Phase1}), (\ref{eq:ET_RLNC_Phase2-user2}), and the right-hand term of (\ref{eq:ET_RLNC_Phase2-user1}).

Under the successive refinement policy, Phase~2 ends once the transmitter receives an ACK from at least one user. Phase~3 then starts, where packets are sent only to the rank-deficient user over a PtP channel. Here, 
two cases arise: (i) user~2 reaches full rank in Phase~2 and sends an ACK while user~1 remains rank-deficient, or (ii) user~1 reaches full rank while user~2 remains rank-deficient. The realized case depends on system parameters, channel conditions, and the chosen $\lambda$. 
Assuming the first case, the transmission time of Phase~2 is given by
$T^{\textnormal{SR-ONOMA}}_{\textnormal{Ph-2}} = T^{\textnormal{ONOMA-(2)}}_{\textnormal{Ph-2}}$,
whose expected value is evaluated in~\eqref{eq:ET_RLNC_Phase2-user2}.
At the end of Phase~2, the transmitter receives an ACK from user~2 and proceeds to transmit the packets of user~1 over a PtP channel.
Accordingly, we derive a lower bound on the expected rank of the decoding matrix
at user~1 at the end of Phase~2 of the SR-ONOMA scheme as 
\begin{equation}
\begin{aligned}
\mathbb{E}\!\big\{N^{\textnormal{SR-ONOMA-(1)}}_{\textnormal{Ph-2}}\big\}
   >&
\bigg(\dfrac{3}{4}\exp\!\Big( 
\dfrac{-\tau(R)\,\sigma^2}{\Omega_{\text{Ph-1}}P}
\bigg)
\\&
+\exp\!\Big(\frac{-\tau(R) \sigma^2}{\Omega_{\text{Ph-2}} P}\Big)\bigg) F\!\Big( \big\lceil \dfrac{\bar{K}}{R} \big\rceil \Big),
\end{aligned}
\label{eq:N_RLNC_Phase2-user1-expanded}
\end{equation}
where
\begin{equation*}
\begin{aligned}
\dfrac{1}{\Omega_{\text{Ph-2}}}=\dfrac{1}{\Omega^{(1)} \lambda'}-\dfrac{1}{(1-\lambda)\Omega^{(2)}}.
\end{aligned}
\end{equation*}The proof is provided in Appendix~\ref{appendix:rank-SR-ONOMA-Ph2}. Once the decoding matrix of user~2 becomes full rank, it returns an ACK to the transmitter, thereby terminating Phase~2 of the transmission.
Subsequently, Phase~3 of the SR-ONOMA scheme begins, during which the packets of user~1 are transmitted over a PtP channel.
Let $T^{\textnormal{SR-ONOMA}}_{\textnormal{Ph-3}}$ denote the transmission time in Phase~3 until the decoding matrix of user~1 attains full rank.
The expected value of $T^{\textnormal{SR-ONOMA}}_{\textnormal{Ph-3}}$ is
upperbounded as
\begin{equation}
\begin{aligned}
\mathbb{E}&\!\big\{ T^{\textnormal{SR-ONOMA}}_{\textnormal{Ph-3}}\big\}\\&
<\exp\!\Big(\frac{\tau(R) \sigma^2}{\Omega^{(1)}P}\Big)\bigg(1-\exp\!\Big( 
\dfrac{-\tau(R)\,\sigma^2}{\Omega_{\text{Ph-1}}P}
\bigg)
\\&
+\exp\!\Big(\frac{-\tau(R) \sigma^2}{\Omega_{\text{Ph-2}} P}\Big) \bigg)F\!\big( \big\lceil \dfrac{\bar{K}}{R} \big\rceil \big).
\end{aligned}
\label{eq:ET_RLNC_Phase3-user1}
\end{equation}
As a result, the total expected completion time of the SR-ONOMA scheme is given by
\begin{equation}
\begin{aligned}
\mathbb{E}&\!\big\{ T^{\textnormal{SR-ONOMA}}\big\}\\&= \mathbb{E}\!\big\{ T^{\textnormal{ONOMA}}_{\textnormal{Ph-1}}\big\}+\mathbb{E}\!\big\{ T^{\textnormal{ONOMA-(1)}}_{\textnormal{Ph-2}}\big\}+\mathbb{E}\!\big\{ T^{\textnormal{SR-ONOMA}}_{\textnormal{Ph-3}}\big\},
\end{aligned}
\label{eq:ET_RLNC_Phase3}
\end{equation}
which is upper bounded by the sum of the right-hand sides of
(\ref{eq:ET_RLNC_Phase1}), (\ref{eq:ET_RLNC_Phase2-user1}), and (\ref{eq:ET_RLNC_Phase3-user1}).

\setcounter{table}{0} 
\renewcommand{\tablename}{Algorithm}
\begin{table}[!t]
\caption{\small{Power allocation policy in ONOMA scheme}}\label{Algorithm:Optimal_power_splitting_ratio}
\vspace{-0.3cm}
\begin{center}
\hspace{0cm}\begin{tabular}{l  }\hline\hline

%{\small{~}}
%{\small{{\small{\textbf{Inputs:} }}}} \\

%{\small{~}}
%{\small{{\small{~$f$ ~~~~~~~~~~~~~~~~~~~~~~~$\%$ Time extension factor }}}} \\
%\hline 

{\small{~}}
{\small{{\small{~$\lambda_u\longleftarrow$ Solution to (\ref{eq:lambda_u})}}}} \\
{\small{~}}
{\small{{\small{~$\lambda_l\longleftarrow$ Solution to (\ref{eq:ET_RLNC_Phase2-user1_lambda_l})}}}} \\

{\small{~}}
{\small{{\small{~If $\lambda_l<\lambda_u$: ~~$\lambda^*=(\lambda_u+\lambda_l)/2$ }}}} \\

{\small{~}}
{\small{{\small{~Else:~~~~~~~~~~~$\lambda^*=\text{RandomChoice}(\lambda_u,\lambda_l)$  }}}} \\

%{\small{~~~}}
%{\small{{\small{~~$\lambda^*=\text{RandomChoice}(\lambda_u,\lambda_l)$ ~~$\%$ for each time slot}}}} \\

\hline
\hline

\end{tabular}
\end{center}
\vspace{-0.6cm}
\end{table}

\subsection{Power allocation policy}
Choosing an efficient power allocation parameter $\lambda$ in Phase~2 is essential to minimize the completion time of the JR- and SR-ONOMA schemes.  Deriving the optimal $\lambda$ is infeasible due to unknown channel gains at the transmitter. Still, at the end of Phase~1, the transmitter can estimate user~2’s channel gain from the time it receives the ACK, denoted by
$T_{\textnormal{Ph-1}}^{\textnormal{ONOMA}} = t_{\textnormal{Ph-1}}^{\textnormal{ONOMA}}$.
Accordingly, the estimated channel gain of user~2, denoted by $\hat{\Omega}^{(2)}$,
is obtained by solving
\begin{equation}
\begin{aligned}
t_\text{Ph-1}^{\text{ONOMA}}=\text{exp}\big( \dfrac{\tau (R) \sigma^2}{\hat\Omega^{(2)}P}\big)F(\big\lceil \dfrac{\bar{K}}{R}\big\rceil).
\end{aligned}
\label{eq:ET_RLNC_Phase1_estimated_gain2}
\end{equation}
Then, to account for the worst case scenario, the channel gain of user 1 is estimated based on the given maximum acceptable transmission time $t_{\text{max}}$ and solving 
\begin{equation}
\begin{aligned}
t_\text{max}=\text{exp}\big( \dfrac{\tau (R) \sigma^2}{\hat\Omega^{(1)}P}\big)F(\big\lceil \dfrac{\bar{K}}{R}\big\rceil).
\end{aligned}
\label{eq:ET_RLNC_Phase1_estimated_gain1}
\end{equation}
Based on the estimated channel gains, we calculate the upper power allocation factor $\lambda_u$ using (\ref{eq:ET_RLNC_Phase2-user2}) by solving 
\begin{equation}
\begin{aligned}
\exp\!\left(\frac{\tau(R) \sigma^2}{(1-\lambda_u)\hat\Omega^{(2)}P}\right)=f\times\exp\!\left(\frac{\tau(R) \sigma^2}{\hat\Omega^{(2)}P}\right),
\end{aligned}
\label{eq:lambda_u}
\end{equation}
where $1<f<3$ denotes the time extension  factor. Then, using (\ref{eq:ET_RLNC_Phase2-user1}), the lower power allocation factor $\lambda_l$ is derived  
\begin{equation}
\begin{aligned}
&\exp\!\Big(\frac{\tau(R) \sigma^2}{\hat\Omega^{(1)} \big(\lambda_l-(1-\lambda_l)\tau(R)\big) P}\Big)
\times \bigg(1-\exp\!\Big( 
\dfrac{-\tau(R)\,\sigma^2}{\hat\Omega_{\text{Ph-1}} P}
\Big)
\bigg)
\\&
=f\times\exp\!\left(\frac{\tau(R) \sigma^2}{\hat\Omega^{(2)}P}\right),
\end{aligned}\label{eq:ET_RLNC_Phase2-user1_lambda_l}
\end{equation}
where
\begin{equation*}
\begin{aligned}
&\dfrac{1}{\hat\Omega_{\text{Ph-1}}}=\dfrac{1}{\hat\Omega^{(1)}} - \dfrac{1}{\hat\Omega^{(2)}}.
\end{aligned}
\end{equation*}
The optimal power allocation factor $\lambda^\ast$ is obtained from the computed bounds $\lambda_l$ and $\lambda_u$, as summarized in Algorithm~\ref{Algorithm:Optimal_power_splitting_ratio}.
In Phase~2, selecting $\lambda < \lambda_u$ maintains a high packet success probability for user~2, while $\lambda > \lambda_l$ ensures reliable decoding for user~1. Hence, if $\lambda_u \ge \lambda_l$, the optimal factor can be set to their mean. Otherwise, when $\lambda_u < \lambda_l$, $\lambda^\ast$ is randomly chosen between $\lambda_u$ and $\lambda_l$ to guarantee sufficient reliability for at least one user at each time slot.

\subsection{Network scheduling}
Consider a network with $U>2$ users, where the transmitter delivers $K$ bits to each user. The RLNC-based ONOMA scheduling extends the two-user ($U=2$) scheme as follows. Packets are first delivered via RLNC-based TDMA to each intended user, while the remaining users overhear and build their decoding matrices. Each user sends an ACK once its matrix reaches full rank $N$. If the intended user has favorable channel conditions, it returns the ACK early, allowing the transmitter to proceed to the next user.

Assume TDMA transmissions for users $u=1$ to $u=i-1$ are completed with early ACKs. The transmitter then serves user $u=i$ while others continue overhearing. If user $u=i$ has a weak channel, its ACK is delayed, whereas another user $u=j$ may respond first. The transmitter thus identifies $u=i$ as weak and $u=j$ as the strongest among the remaining users, and initiates Phase~2 ONOMA with symbol-aware NOMA for users $u=i$ and $u=j$, while others decode packets for $u=i$. As stronger users reach full rank, NOMA pairing is updated until user $u=i$ sends an ACK, after which TDMA resumes.

\begin{figure*}
    \centering
    \includegraphics[width=1\linewidth]{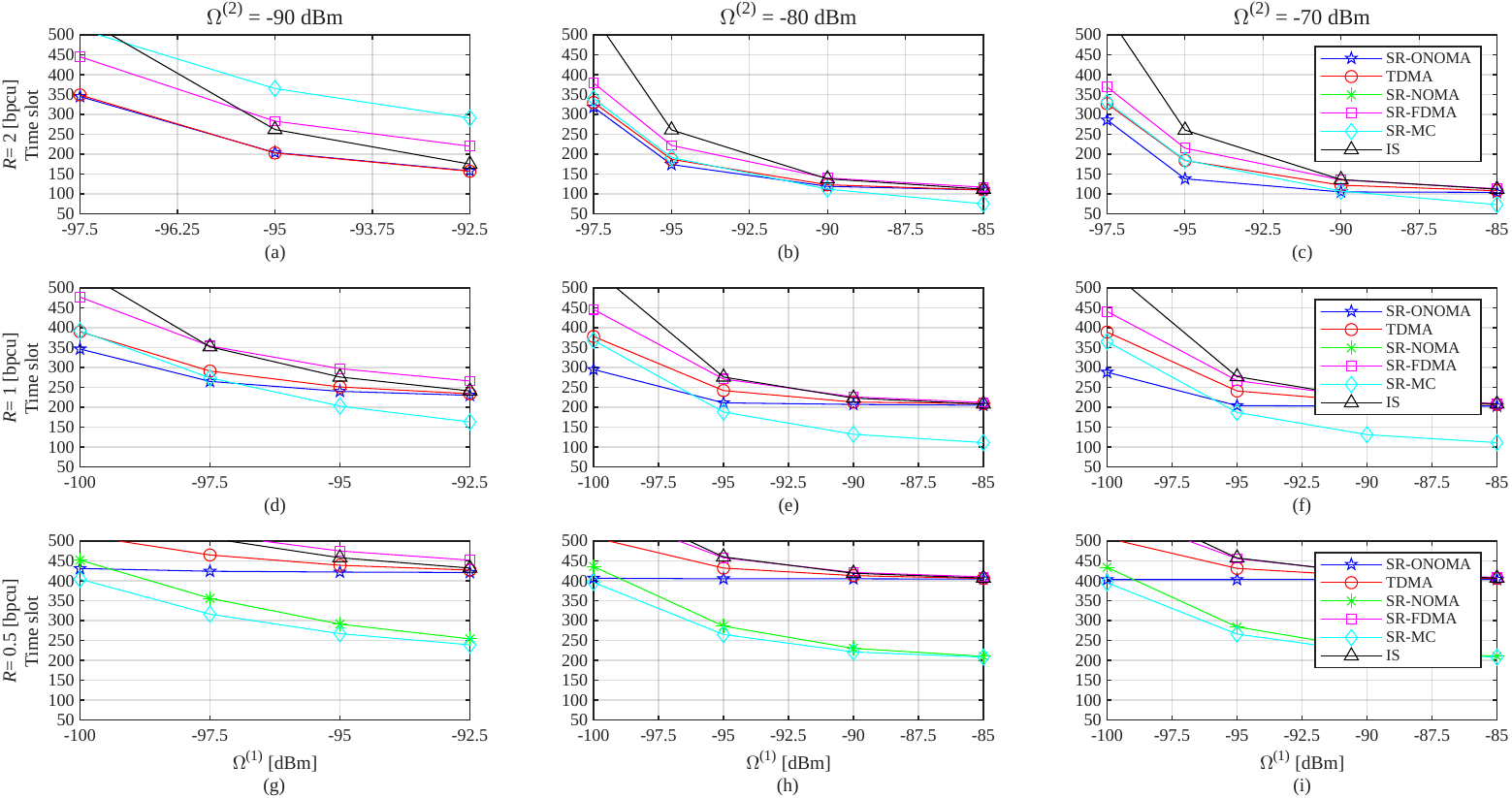}
    \vspace{-0.7cm}
    \caption{Comparison of the expected completion time of the SR-ONOMA scheme with the baseline transmission schemes in a downlink two-user low-feedback network with block of $K = N R n_{\text{FFT}} = 102.4\,\text{kbits}$ data bits for each user, IFFT size $n_{\text{FFT}}=1024$, symbol time $T_{\text{sym}}=6.67$ \textmu s in  one packet per OFDM symbol  policy; a,b,c) Data rate $R=2$ bpcu and $N=5$ packets per data block, suitable for networks with a minimum acknowledgment interval $T_{\text{ACK}}<N\times T_{\text{sym}}= 333.7$ \textmu s, with the channel gain of user 2 set to $\Omega^{(2)}=-90, -80$, and $-70$\,dBm, respectively. d,e,f) Data rate $R=1$ bpcu and $N=100$ packets per data block, suitable for networks with a minimum acknowledgment interval $T_{\text{ACK}}<N\times T_{\text{sym}}= 667$ \textmu s, with the channel gain of user 2 set to $\Omega^{(2)}=-90, -80$, and $-70$\,dBm, respectively. g,h,i) Data rate $R=0.5$ bpcu and $N=200$ packets per data block, suitable for networks with a minimum acknowledgment interval $T_{\text{ACK}}<N\times T_{\text{sym}}= 1334$ \textmu s, with the channel gain of user 2 set to $\Omega^{(2)}=-90, -80$, and $-70$\,dBm, respectively.
    }
    \vspace{-0.2cm}
    \label{fig:simulation}
\end{figure*}

\section{Simulation Results}
We evaluate the RLNC-based ONOMA scheme against baseline methods in terms of completion time in a two-user low-feedback network. Initially, the transmitter has no channel state information or statistical knowledge. For fairness, the SR transmission policy is applied to both proposed and baseline schemes. The expected completion time is obtained by averaging over 300 independent simulation runs.

Fig.~\ref{fig:simulation} shows the average completion time (average number of packet transmissions until both users reach full rank) for the proposed and baseline schemes under different data rates, packet numbers, and channel gains. Each packet is sent in one OFDM symbol. The subcarrier spacing is $B_{\text{sub}}=15$~kHz, giving $T_{\text{sym}}=6.67~\text{\textmu s}$, and the total bandwidth $B=15.36$~MHz yields $n_{\text{FFT}}=1024$ subcarriers, all used for data.
The number of bits per block per user is $K=NRn_{\text{FFT}}=102.4$~kbits, with $\bar K=NR=100$ in all scenarios. For SR-ONOMA, the time extension factor is set to $f=1.5$.

In Fig.~\ref{fig:simulation}a--c), the  rates are $R^{(1)}=R^{(2)}=R=2$\,bpcu, giving $N^{(1)}=N^{(2)}=N=\bar K/R=50$ packets. Thus, ACKs must be sent within $T_{\text{ACK}}=N T_{\text{sym}}=333.5$~\textmu s. The IS scheme requires rank $2N$, forcing continued transmission until both ACKs are received, which increases latency under weak channels. In SR-FDMA, two  successful packets are needed per rank increment in Phase~1, further prolonging completion time. For SR-NOMA with $\lambda=0.5$, the condition  \eqref{eq:lambda-requirement} is not satisfied, leading to infinite packet transmissions.
As shown in Fig.~\ref{fig:simulation}a), for $\Omega^{(2)}=-90$~dBm, SR-ONOMA and TDMA achieve the smallest completion time for all $\Omega^{(1)}$. Fig.~\ref{fig:simulation}c) shows that when user~2 is strong and user~1 is weak, SR-ONOMA outperforms TDMA by up to $34\%$. When both channels are strong, most schemes perform similarly, except SR-MC, which achieves lower latency by combining both users’ packets in Phase~1. However, since each user must decode at rate $2R=4$\,bpcu, SR-MC suffers sharp latency increases under weak channels due to frequent packet losses.

In Fig.~\ref{fig:simulation}d--f), $R^{(1)}=R^{(2)}=R=1$\,bpcu, giving $N^{(1)}=N^{(2)}=N=\bar K/R=100$ packets. Hence, ACKs must be delivered within $T_{\text{ACK}}=N T_{\text{sym}}=667$~\textmu s. As before, IS and SR-FDMA exhibit large completion times, especially when one channel is weak. For SR-NOMA with $\lambda=0.5$, the condition \eqref{eq:lambda-requirement} is not satisfied, leading to infinite transmissions.
Under strong channels for both users, SR-MC outperforms others since both can decode at $2R=2$\,bpcu. However, when user~1 has a weak channel, SR-ONOMA surpasses the second-best scheme (SR-MC) by up to $27\%$ at $\Omega^{(1)}=-100$~dBm.

\begin{figure*}
    \centering
    \includegraphics[width=1\linewidth]{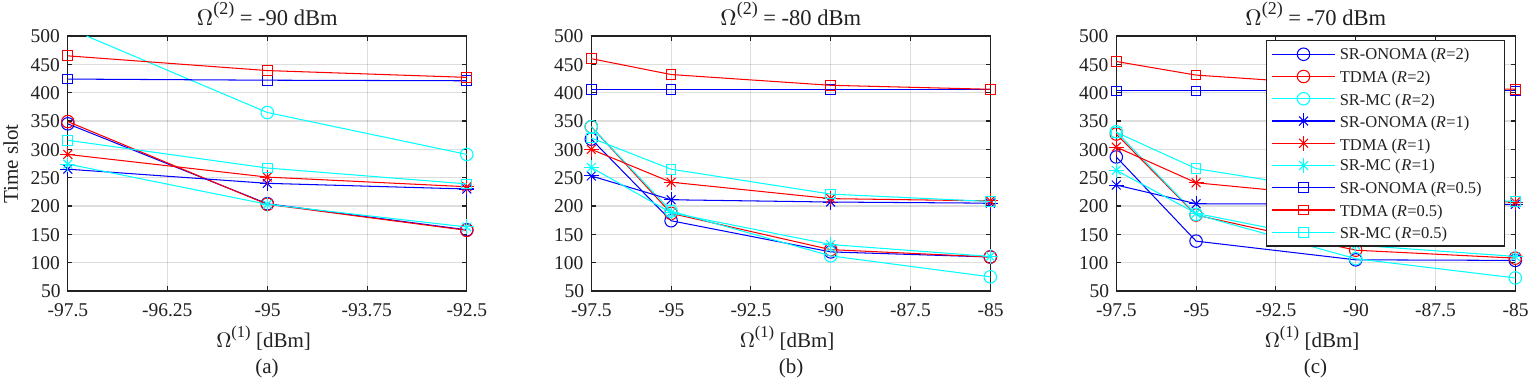}
    \vspace{-0.7cm}
    \caption{Comparison of the expected completion time of the SR-ONOMA scheme with the most competitive baseline schemes, including TDMA and SR-MC, as illustrated in Fig.~\ref{fig:simulation}, for the selected values of $R=2,1$ and $0.5$ bpcu with; a)   $\Omega^{(2)}=-90$ dBm, b) $\Omega^{(2)}=-80$ dBm, and c) $\Omega^{(2)}=-70$\,dBm.}
    \vspace{-0.5cm}
    \label{fig:simulation-final}
\end{figure*}

In Fig.~\ref{fig:simulation}g--i), $R^{(1)}=R^{(2)}=R=0.5$\,bpcu, yielding $N^{(1)}=N^{(2)}=N=\bar K/R=200$ packets. Thus, ACKs must be delivered within $T_{\text{ACK}}=N T_{\text{sym}}=1334$~\textmu s.
In this regime, SR-ONOMA outperforms TDMA, IS, and SR-FDMA, but SR-MC achieves the smallest completion time due to the lower rate. A similar gain is observed for SR-NOMA. This is because the condition \eqref{eq:lambda-requirement} holds for $\lambda=0.5$, allowing both users to decode packets for both users during Phase~1.

Fig.~\ref{fig:simulation-final} summarizes the results by comparing SR-ONOMA, TDMA, and SR-MC at $R=2$, $1$, and $0.5$\,bpcu. In Fig.~\ref{fig:simulation-final}b)–c), when user~2 has $\Omega^{(2)}=-80$ or $-70$~dBm, all schemes achieve the smallest completion time at $R=2$\,bpcu and the largest at $R=0.5$\,bpcu.
However, for a severely degraded channel $\Omega^{(2)}=-90$~dBm, the completion time at $R=2$,bpcu increases significantly, as shown in Fig.~\ref{fig:simulation-final}a). Moreover, when user~2’s channel is strong, SR-ONOMA at $R=2$ and $R=1$\,bpcu achieves the lowest completion time, especially when user~1’s channel is weak.

%%%%%%%%%%%%%%%%%%%
Consequently, in networks where user acknowledgments are scarce, the SR-ONOMA scheme outperforms all baseline transmission methods in terms of completion time, particularly under highly asymmetric channel conditions and in high data-rate regimes. This performance advantage primarily arises from the joint exploitation of implicit channel-gain estimation and symbol-aware NOMA transmission inherent to the SR-ONOMA framework.  

\section{Conclusion}
This paper proposed RLNC-based Overhearing-Driven NOMA (ONOMA) for minimizing completion latency in low-feedback multi-user networks with unknown and asymmetric channels. Unlike conventional schemes requiring explicit CSI or frequent feedback, ONOMA exploits sparse ACK timing as an implicit estimator of channel statistics. The early acknowledgment from overhearing users reveals relative channel strength, enabling adaptive power allocation without instantaneous CSI.
A second key contribution is reconstruction-based symbol cancellation. After decoding the weak user’s RLNC packets during the overhearing phase, the strong user regenerates the weak user’s encoded symbols in the NOMA phase and performs exact symbol-level subtraction, eliminating interference without classical decoding-based SIC.
By combining ACK-timing–based channel inference with symbol-aware interference mitigation, ONOMA decouples user latencies and mitigates weakest-user domination, achieving substantial completion-time reductions under severe channel asymmetry.

\appendices
\section{Expected rank of decoding matrix of the weaker user at the end of Phase 1 of SR-MC transmission} \label{appendix:rank-SR-MC}
 From \eqref{eq:ET_RLNC}, the expected time so a user reaches the rank $k<N$ is given by 
\begin{equation}
\begin{aligned}
\mathbb{E}&\{T|k\}=\sum_{r=0}^{k-1}\dfrac{1}{s_r}=\dfrac{1}{P_{\text{succ}}}\sum_{r=0}^{k-1}\dfrac{1}{1-2^{r-N}}.
\end{aligned}
\label{eq:ET_RLNC_rank_k-multicast-appendix}
\end{equation}
So, denoting the transmission time at Phase 1 and the rank of decoding matrix at user 1 at the end of Phase 1 by $T_{\text{Ph-1}}^{\text{SR-MC}}$ and $N_{\text{Ph-1}}^{\text{SR-MC-(1)}}$, respectively, (\ref{eq:ET_RLNC_rank_k-multicast-appendix}) is rewritten as 
\begin{equation}
\begin{aligned}
 &P_{\text{succ, Ph-1}}^{\text{SR-MC-(1)}}\,\mathbb{E}\{T_{\text{Ph-1}}^{\text{SR-MC}}|N_{\text{Ph-1}}^{\text{SR-MC-(1)}}\}\\& =\sum_{r=0}^{N^{\text{SR-MC-(1)}}_\text{Ph-1}-1}\dfrac{1}{1-2^{r-N^{(1)}}}< S(N^{\text{SR-MC-(1)}}_\text{Ph-1})N^{\text{SR-MC-(1)}}_\text{Ph-1},\\& ~~~~~~~~~~~~~~~~~~~~~~~~~~~~~~~\leq  \dfrac{4}{3}N^{\text{SR-MC-(1)}}_\text{Ph-1},
\end{aligned}
\label{eq:ET_RLNC_SR_MC_Ph_1}
\end{equation}
where 
\begin{equation*}
\begin{aligned}
  S(x)= \dfrac{1}{1-2^{x-N^{\text{(1)}}-1}}.
\end{aligned}
\end{equation*}
Here, $P_{\text{succ, Ph-1}}^{\text{SR-MC-(1)}}$ denotes the packet success probability at user~1 during Phase~1, and the first inequality follows from the fact that each term in the summation is greater than one, with the largest term being $S(N^{\text{SR-MC-(1)}}_\text{Ph-1})$.
Further, since  $N^{\text{SR-MC-(1)}}_\text{Ph-1}<N^{(1)}$, we have
$S(N^{\text{SR-MC-(1)}}_\text{Ph-1})\leq S(N^{(1)}-1)=4/3$, which justifies the second inequality in \eqref{eq:ET_RLNC_SR_MC_Ph_1}.
 Therefore, by applying the tower property, we take the expectation of both sides of \eqref{eq:ET_RLNC_SR_MC_Ph_1} with respect to $N^{\text{SR-MC-(1)}}_{\text{Ph-1}}$, which yields
\begin{equation}
\begin{aligned}
\mathbb{E}&\{N^{\text{SR-MC-(1)}}_\text{Ph-1}\}>\dfrac{3}{4}P_{\text{succ, Ph-1}}^{\text{SR-MC-(1)}}\mathbb{E}\{T_\text{Ph-1}^{\text{SR-MC}}\}.
\end{aligned}
\label{eq:EN_RLNC_SR-MC-Phase1_appendix}
\end{equation}%wherein the  expectation of  $T_{\text{Ph-1}}^{\text{SR-MC}}$ is described in (\ref{eq:ET_SR-multicast-Ph1}).
Also, the packet success probability of user 1 during Phase 1 is formulated as
\begin{equation}
\begin{aligned}
P_{\text{succ, Ph-1}}^{\text{SR-MC-(1)}}=\exp(\dfrac{-\tau(R)\sigma^2}{\Omega^{(1)}P}).
\end{aligned}
\label{eq:P_loss1-SR-MC_Phase1}
\end{equation} 
Finally, by inserting (\ref{eq:P_loss1-SR-MC_Phase1}) and (\ref{eq:ET_SR-multicast-Ph1}) into (\ref{eq:EN_RLNC_SR-MC-Phase1_appendix}), the lowerbound to the expected rank of the decoding matrix of user 1 at the end of Phase 1 of SR-MC scheme is written as
\begin{equation}
\begin{aligned}
\mathbb{E}&\{N^{\text{SR-MC-(1)}}_\text{Ph-1}\}> \dfrac{3}{4}\exp\big(\dfrac{-\tau(R)\sigma^2}{\Omega^{(1)}P}\big)\text{exp}\big( \dfrac{\tau (R) \sigma^2}{\Omega^{(2)}P}\big)F(\big\lceil 2\dfrac{\bar{K}}{R}\big\rceil),
\end{aligned}
\label{eq:EN_RLNC_Phase1-expanded-appendix}
\end{equation} and therefore, (\ref{eq:EN_RLNC_SR-MC-Phase1}) follows.

\section{Expected time of the Phase 2 of SR-MC scheme}\label{Appendix:ET-phase2}
At the beginning of the Phase 2 of the transmission, the  rank of the decoding matrix at user 1 is $N^{\textnormal{SR-MC-(1)}}_{\textnormal{Ph-1}}$, which is smaller than $N^{(1)}=\big\lceil 2\bar{K}/R\big\rceil$. From (\ref{eq:ET_RLNC}) with a given $N^{\textnormal{SR-MC-(1)}}_{\textnormal{Ph-1}}$, the expected time required for the decoding matrix of user 1 to reach the rank of $N^{(1)}$ at Phase 2 is written as 
\begin{equation}
\begin{aligned}
&P_{\text{succ, Ph-2}}^{\text{SR-MC-(1)}}\,\mathbb{E}\!\big\{T^{\textnormal{SR-MC}}_{\textnormal{Ph-2}}|N^{\textnormal{SR-MC-(1)}}_{\textnormal{Ph-1}}\big\} \\& =\sum_{r= N_{\text{Ph-1}}^{\text{SR-MC-(1)}}}^{N^{(1)}-1}\dfrac{1}{1-2^{r-N^{(1)}}}
\\&
= \Big(\sum_{r=0}^{N^{(1)}-1}\dfrac{1}{1-2^{r-N^{(1)}}}-\sum_{r=0}^{N_{\text{Ph-1}}^{\text{SR-MC-(1)}}-1}\dfrac{1}{1-2^{r-N^{(1)}}}\Big).
\end{aligned}
\label{eq:ET_RLNC_SR-MC-Phase2-appendix}
\end{equation} The first summation term in (\ref{eq:ET_RLNC_SR-MC-Phase2-appendix}) is $F(N^{(1)})=F\big(\big \lceil \dfrac{2\bar K}{R}\big \rceil\big)$, while the elements of the second summation term are larger than 1. Therefore, from (\ref{eq:ET_RLNC_SR-MC-Phase2-appendix}), the conditional transmission time at Phase 2 is rewritten as
\begin{equation}
\begin{aligned}
\mathbb{E}\!\big\{T^{\textnormal{SR-MC}}_{\textnormal{Ph-2}}|N^{\textnormal{SR-MC-(1)}}_{\textnormal{Ph-1}}\big\} <\dfrac{1}{P_{\text{succ, Ph-2}}^{\text{SR-MC-(1)}}} \Big(F\!\big( \big\lceil \dfrac{2\bar{K}}{R} \big\rceil \big)-N_{\text{Ph-1}}^{\text{SR-MC-(1)}}\Big).
\end{aligned}
\label{eq:ET_RLNC_SR-MC-Phase2-appendix2}
\end{equation}
Here, since the transmitter encodes only  the packets of user 1, this user only requires to decode the received packets with rate $R/2$. Therefore, the packet success probability at user 1 during Phase 2 of SR-MC transmission scheme is written as    
\begin{equation}
\begin{aligned}
P_{\text{succ, Ph-2}}^{\text{SR-MC-(1)}}=\exp\Big(\frac{-\tau(R/2) \sigma^2}{\Omega^{(1)} P}\Big).
\end{aligned}
\label{eq:P_succ_SR-MC-phase2-appendix}
\end{equation}
So,  by taking the expectation from the both sides of (\ref{eq:ET_RLNC_SR-MC-Phase2-appendix2}) in terms of $N^{\text{SR-MC-(1)}}_\text{Ph-1}$, we have
\begin{equation}
\begin{aligned}
& \mathbb{E}\!\big\{T^{\textnormal{SR-MC}}_{\textnormal{Ph-2}}\big\}<\dfrac{1}{P_{\text{succ, Ph-2}}^{\text{SR-MC-(1)}}} \Big(F\!\big( \big\lceil \dfrac{2\bar{K}}{R} \big\rceil \big)-\mathbb{E\{}N^{\textnormal{SR-MC-(1)}}_{\textnormal{Ph-1}}\}\Big)\\& <\exp\Big(\frac{\tau(R/2) \sigma^2}{\Omega^{(1)} P}\Big)\\&
\times \bigg(1-\dfrac{3}{4}\exp\!\Big( 
\dfrac{-\tau(R)\,\sigma^2}{P}
\big( \dfrac{1}{\Omega^{(1)}} - \dfrac{1}{\Omega^{(2)}} \big)
\Big)\bigg)
F\!\Big( \big\lceil \dfrac{2\bar{K}}{R} \big\rceil \Big),
\end{aligned}
\label{eq:ET_RLNC_Phase2-appendix3}
\end{equation} where the first inequality is stem by applying the tower property, and the second inequality is derived from (\ref{eq:P_succ_SR-MC-phase2-appendix}) and (\ref{eq:EN_RLNC_SR-MC-Phase1}).  Therefore, \eqref{eq:ET_RLNC_SR-MC_Phase2-user1} follows.

\section{Expected rank of decoding matrix of the weaker user at the end of Phase 2 of SR-ONOMA scheme} \label{appendix:rank-SR-ONOMA-Ph2}
Similar to the arguments in Appendix A, we have
\begin{equation}
\begin{aligned}
\mathbb{E}\{N^{\text{SR-ONOMA-(1)}}_\text{Ph-2}\}&>\mathbb{E}\{N^{\text{ONOMA-(1)}}_\text{Ph-1}\}\\&+P_{\text{succ, Ph-2}}^{\text{ONOMA-(1)}}  \mathbb{E}\{T_\text{Ph-2}^{\text{SR-ONOMA-(2)}}\},
\end{aligned}
\label{eq:EN_RLNC_SR-ONOMA-Phase2}
\end{equation}where the lower bound on the expected value of
$N_{\textnormal{Ph-1}}^{\textnormal{ONOMA-(1)}}$ is given in
\eqref{eq:EN_RLNC_Phase1-expanded},
the packet success probability of user~1 in Phase~2 is provided in
\eqref{eq:P_loss_Phase2-1-user-1},
and the expected time for the decoding matrix of user~2 to reach full rank
during Phase~2 of the transmission is derived in
\eqref{eq:ET_RLNC_Phase2-user2}.
Therefore, by substituting
\eqref{eq:EN_RLNC_Phase1-expanded},
\eqref{eq:P_loss_Phase2-1-user-1}, and
\eqref{eq:ET_RLNC_Phase2-user2}
into \eqref{eq:EN_RLNC_SR-ONOMA-Phase2},
the result in \eqref{eq:N_RLNC_Phase2-user1-expanded} follows.

\section{Expected time of the Phase 3 of SR-ONOMA}
During Phase~3, the packets of user~1 are transmitted over a PtP channel until the decoding matrix of user~1 attains full rank.
Given this rank at the beginning of Phase~3,
the expected completion time of Phase~3
is upper bounded as
\begin{equation}
\begin{aligned}
\mathbb{E}&\!\big\{ T^{\textnormal{SR-ONOMA}}_{\textnormal{Ph-3}}|N^{\textnormal{SR-ONOMA-(1)}}_{\textnormal{Ph-2}}\big\}\\&=\dfrac{1}{P_{\text{succ}}^{\text{PtP-(1)}}}\sum_{r=N^{\textnormal{SR-ONOMA-(1)}}_{\textnormal{Ph-2}}}^{N-1}\dfrac{1}{1-2^{r-N}}\\&=\dfrac{1}{P_{\text{succ}}^{\text{PtP-(1)}}}\bigg(\sum_{r=0}^{N-1}\dfrac{1}{1-2^{r-N}}-\sum_{r=0}^{N^{\textnormal{SR-ONOMA-(1)}}_{\textnormal{Ph-2}}-1}\dfrac{1}{1-2^{r-N}}\bigg)\\ &<\exp\!\Big(\frac{\tau(R) \sigma^2}{\Omega^{(1)}P}\Big) \Big(F\!\big( \big\lceil \dfrac{\bar{K}}{R} \big\rceil \big)-N^{\textnormal{SR-ONOMA-(1)}}_{\textnormal{Ph-2}}\Big),
\end{aligned}
\label{eq:ET_RLNC_Phase3-user1-conditional}
\end{equation} where $P_{\text{succ}}^{\text{PtP-(1)}}=P_{\text{succ,Ph-1}}^{\text{ONOMA-(1)}}$ is derived from \eqref{eq:P_loss_RLNC_TDMA} and 
the inequality is derived from the fact that the elements of the summation term are larger than 1. Then, taking expectation from (\ref{eq:ET_RLNC_Phase3-user1-conditional}), we have

\begin{equation}
\begin{aligned}
\mathbb{E}&\!\big\{ T^{\textnormal{SR-ONOMA}}_{\textnormal{Ph-3}}\big\}\\ &<\exp\!\Big(\frac{\tau(R) \sigma^2}{\Omega^{(1)}P}\Big) \Big(F\!\big( \big\lceil \dfrac{\bar{K}}{R} \big\rceil \big)-\mathbb{E}\big\{N^{\textnormal{SR-ONOMA-(1)}}_{\textnormal{Ph-2}}\big\}\Big)\\&
<\exp\!\Big(\frac{\tau(R) \sigma^2}{\Omega^{(1)}P}\Big)\bigg(1-\dfrac{3}{4}\exp\!\Big( 
\dfrac{-\tau(R)\,\sigma^2}{\Omega_{\text{Ph-1}}P}
\bigg)
\\&
-\exp\!\Big(\frac{-\tau(R) \sigma^2}{\Omega_{\text{Ph-2}} P}\Big) \bigg)F\!\big( \big\lceil \dfrac{\bar{K}}{R} \big\rceil \big),
\end{aligned}
\label{eq:ET_RLNC_Phase3-user1-appendix}
\end{equation} where \eqref{eq:N_RLNC_Phase2-user1-expanded} is used. Thus, from \eqref{eq:ET_RLNC_Phase3-user1-appendix}, (\ref{eq:ET_RLNC_Phase3-user1}) is followed.

%\section*{Acknowledgment}



\ifCLASSOPTIONcaptionsoff
  \newpage
\fi

\bibliographystyle{IEEEtran}

\bibliography{references.bib}

\end{document}